\documentclass{article} 
\usepackage{iclr2024_conference,times}

\usepackage{amsmath,amsfonts,bm}

\def\eqref#1{equation~\ref{#1}}

\def\1{\bm{1}}

\DeclareMathAlphabet{\mathsfit}{\encodingdefault}{\sfdefault}{m}{sl}
\SetMathAlphabet{\mathsfit}{bold}{\encodingdefault}{\sfdefault}{bx}{n}

\DeclareMathOperator*{\argmax}{arg\,max}
\DeclareMathOperator*{\argmin}{arg\,min}

\usepackage{hyperref}
\usepackage{url}

\usepackage{multirow}

\usepackage{setspace}
\usepackage{wrapfig}
\usepackage{comment}
\usepackage{csquotes}
\usepackage{subcaption}
\usepackage{enumitem}
\usepackage{adjustbox}

\usepackage{amsmath}
\usepackage{amssymb}
\usepackage{mathtools}
\usepackage{amsthm}
\usepackage{algorithm2e}
\usepackage{xcolor}

\theoremstyle{plain}

\theoremstyle{definition}

\theoremstyle{remark}

\newcommand{\ourmodel}{{\small\textsf{SAPA}}\xspace}

\title{Sharpness-Aware Data Poisoning Attack}

\author{Pengfei He$^1$, Han Xu$^1$, Jie Ren$^1$, Yingqian Cui$^1$, Shenglai Zeng$^1$, Hui Liu$^1$,\\ \textbf{Charu C. Aggarwal$^2$, Jiliang Tang$^1$} \\
$^1$Department of Computer Science and Engineering, Michigan State University\\
$^2$IBM T. J. Watson Research Center, New York\\
$^1$\texttt{\{hepengf1,xuhan1,renjie3,cuiyingq,zengshe1,liuhui7,tangjili\}@msu.edu}\\
$^2$\texttt{charu@us.ibm.com}
}

\newcommand{\blue}[1]{\textcolor{violet}{\textbf{#1}}}

\iclrfinalcopy 
\begin{document}

\maketitle

\begin{abstract}
Recent research has highlighted the vulnerability of Deep Neural Networks (DNNs) against data poisoning attacks. These attacks aim to inject poisoning samples into the models' training dataset such that the trained models have inference failures. While previous studies have executed different types of attacks, one major challenge that greatly limits their effectiveness is the uncertainty of the re-training process after the injection of poisoning samples. It includes the uncer-
tainty of training initialization, algorithm and model architecture. To address this challenge, we propose a new strategy called ``\textit{Sharpness-Aware Data Poisoning Attack (\ourmodel)}''. In particular, it leverages the concept of DNNs' loss landscape sharpness to optimize the poisoning effect on the (approximately) worst re-trained model. 
Extensive experiments demonstrate that \ourmodel offers a general and principled strategy that significantly enhances numerous poisoning attacks against various types of re-training uncertainty.
\end{abstract}

\vspace{-0.5cm}
\section{Introduction}\label{sec:intro}
\vspace{-0.3cm}

The rise of machine learning (ML) models that collect training data from public sources, such as large language models~\citep{brown2020language, radford2019language} and large visual models~\citep{ramesh2021zero, radford2021learning}, highlights the need of robustness against data poisoning attacks~\citep{steinhardt2017certified, shafahi2018poison, chen2017targeted}. Data poisoning attack refers to the threats of an adversary injecting poisoning data samples into the collected training dataset, such that the trained ML models can have malicious behaviors. For example, by injecting poisoning samples, the adversary's objective is to cause a poisoned model has a poor overall accuracy (known as \textit{un-targeted attacks}~\citep{steinhardt2017certified, li2020gradient, ren2022transferable}), or misclassifies a specified subset of test samples (known as \textit{targeted attacks}~\citep{shafahi2018poison, zhu2019transferable}).
Additionally, in \textit{backdoor attacks}~\citep{chen2017targeted, saha2020hidden, tran2018spectral}, the adversary aims to create ``backdoors'' in the poisoned model such that the model gives a specific output as the adversary desires if the backdoor trigger is presented, regardless of the actual input. 

Many poisoning attacks in Deep Neural Networks (DNNs) face a common obstacle that limits their effectiveness, which is the uncertainty of the re-training process after the injection of poisoning samples.
This challenge is also highlighted in previous studies such as ~\citep{schwarzschild2021just, zhu2019transferable, huang2020metapoison, geiping2020witches}. In particular, most  existing methods (as reviewed in Section~\ref{sec:related} and Appendix~\ref{app: review}) generate poisoning samples based on the effect of only one victim model~\citep{shafahi2018poison, souri2022sleeper}. However, on the poisoned dataset, the re-trained model may converge to a very different point due to the re-training uncertainty, such as training algorithm, model initialization and model architectures. As a consequence, the injected poisoning samples could lose their efficacy and the poisoning effect is compromised. To overcome this difficulty, recent methods~\citep{geiping2020witches,huang2020metapoison} have introduced the \textit{Ensemble and Re-initialization (E\&R) Strategy:} which proposes to optimize the average poisoning effect of multiple models with various model-architectures and initializations. However, this method tends to be computationally inefficient~\citep{huang2020metapoison}. 
In general, if we consider the poisoning attacks as general bilevel optimization problems~\citep{bard2013practical, colson2007overview} (Eq.~\ref{eq: general poison} in Section~\ref{sec:definition}), poisoning problems in DNNs can be categorized as ``multiple inner minima'' bilevel optimization problems~\citep{sow2022primaldual, liu2020generic, li2020improved}.  
Prior works~\citep{sow2022primaldual, li2020improved, liu2020generic, liu2021towards} have demonstrated the theoretical difficulty and complexity of solving ``multiple inner minima'' problems. 

In this work, we introduce a new attack strategy, \textit{\textbf{S}harpness-\textbf{A}ware Data \textbf{P}oisoning \textbf{A}ttack (\ourmodel)}. 
In this method, we aim to inject poisoning samples, which \textbf{optimizes the poisoning effect of the approximately ``worst'' re-trained model}. In other words, even the worst re-trained model (which achieves the relatively worst poisoning effect) can have a strong poisoning effect. 
Furthermore, we find that this strategy can be successfully accomplished by tackling the victim model's 
loss landscape sharpness~\citep{foret2020sharpness}. Notably, the loss landscape sharpness is more frequently used to explain the generalization characteristic of DNNs. In our work, we show the possibility to leverage the algorithms of sharpness (such as~\citep{foret2020sharpness}) to advance poisoning attacks. Through experimental studies, we show that our proposed method \textbf{\ourmodel is a general and principle strategy} and it can improve the performance of various poisoning attacks, including targeted attacks, un-targeted attacks and backdoor attacks (Section~\ref{exp:targeted}, \ref{exp:backdoor}\&~\ref{exp:untargeted}).  Moreover, we show \ourmodel can also effectively improve the poisoning effect under various re-training uncertainties, including for re-training algorithm as well as model architectures (Section~\ref{sec:exp3}). Compared to the E\&R strategy~\citep{huang2020metapoison}, we show \ourmodel is more computational efficient (Section~\ref{sec:exp2}). 

\vspace{-0.3cm}
\section{Related work}\label{sec:related}
\vspace{-0.2cm}

\subsection{End-to-end Data Poisoning Attacks}
\vspace{-0.3cm}

Data poisoning attacks~\citep{biggio2012poisoning,steinhardt2017certified} refer to the adversarial threat during the data collection phase of training ML models. These attacks manipulate the training data so that the trained models have malicious behaviors. Common objectives of poisoning attacks include the purposes of causing a poisoned model to have a poor overall accuracy (\textit{un-targeted attacks}~\citep{steinhardt2017certified, li2020gradient, ren2022transferable}), misclassifying a specified subset of test samples (\textit{targeted attacks}~\citep{shafahi2018poison, zhu2019transferable}), or insert backdoors (\textit{backdoor attacks}~\citep{chen2017targeted, saha2020hidden, tran2018spectral}). Notably, in this work, we focus on the poisoning attacks in an \textit{\textbf{``end-to-end''}} manner, which means that the victim model is trained on the poisoned dataset from scratch. It is a practical setting for poisoning attacks as the attackers cannot take control of the re-training process. 

In general, the feasibility of poisoning attacks highly depends on the complexity of the victim model. For linear models, such as logistic regression and SVM, there are methods~\citep{biggio2012poisoning,steinhardt2017certified, koh2022stronger} that can find solutions close to optimal ones. For DNNs, the exact solution is generally considered intractable, due to the complexity and uncertainty of the re-training process. In detail, in the end-to-end scenario, the victim model can be trained with a great variety of choices, including various choices of model architectures, initialization, and training algorithms, which are hard to thoroughly consider while generating the poisoning samples. Faced with this difficulty, many works~\citep{shaban2019truncated} focus on the transfer learning setting~\citep{shafahi2018poison, zhu2019transferable}, where they loosen the ``end-to-end'' assumption. They assume the attacker has knowledge of a pre-trained model, and the victim model is fine-tuned on this pre-trained model. However, these attacks are usually ineffective in the end-to-end setting. Recent works such as Gradient Matching~\citep{geiping2020witches} make a great progress under end-to-end setting. However, these attacks still face obvious degradation when there are re-training variations such as model architectures.
To further improve the attacks, \citet{huang2020metapoison} devise the ``ensemble and re-initialization'' strategy to take several victim models into consideration. However, it is time and memory-consuming, and cannot sufficiently cover the re-training possibilities. In Appendix~\ref{app: review}, we provide a more comprehensive review of existing poisoning attacks. In Appendix \ref{app:b2}, we introduce some poisoning attacks which consider the scenario different from end-to-end scenario.



\vspace{-0.3cm}
\subsection{Loss Landscape Sharpness}
\vspace{-0.3cm}
In this paper, our proposed method involves calculating and optimizing the loss landscape sharpness of DNNs. 
The notion of the loss landscape sharpness and its connection to generalization has been extensively studied, both empirically \citep{keskar2016large, dinh2017sharp} and theoretically~\citep{dziugaite2017computing, neyshabur2017pac}. These studies have motivated the development of methods \citep{izmailov2018averaging, chaudhari2019entropy} that aim to improve model generalization by manipulating or penalizing sharpness. Among these methods, Sharpness-Aware Minimization (SAM)~\citep{foret2020sharpness} has shown to be highly effective and scalable for DNNs. In this paper, we explore the use of sharpness for data poisoning attacks.
\vspace{-0.3cm}
\section{Preliminary}\label{sec:definition}
\vspace{-0.2cm}

In this section, we introduce the definitions of the (loss landscape) sharpness, as well as the formulations of several most frequently studied data poisoning attacks. 
We start by providing some necessary notations. In this paper, we focus on classification tasks, with input $x\in \mathcal{X}$ and label $y\in \mathcal{Y}$ following the distribution $D$ which is supported on $\mathcal{X}\times \mathcal{Y}$. Under this dataset, a classification model $f(\cdot;\theta):\mathcal{X}\rightarrow \mathcal{Y}$ is trained on the training set $D_{tr}=\{(x_i,y_i), i = 1,...,n\}$, whose $n$ data samples follow $D$. The model parameter $\theta$ is from the parameter space $\Theta$.
We define the loss function as $l(f(x;\theta),y)$, and the   (training) loss as $L(\theta;D_{tr})=\frac{1}{n}\sum_{i=1}^nl(f(x_i;\theta),y_i)$.

\vspace{-0.2cm}
\subsection{Loss Landscape Sharpness}
\vspace{-0.1cm}
We follow the precedent work~\citep{foret2020sharpness} to define the loss landscape sharpness (refer to as ``sharpness'' for simplicity) as in Eq.~\ref{eq:sharpness}. It measures how quickly the model's training loss can be increased by moving its parameter to a nearby region.  Following Eq.~\ref{eq:sharpness}, it calculates the training loss increase after the model parameter $\theta$ is perturbed by $v$, whose $l_p$ norm is constrained by $||v||_p\leq \rho$:
\vspace{-0.1cm}
\begin{align}\label{eq:sharpness}\small
\begin{split}
   S^{\rho}(\theta; D_{tr}) & = \max_{||v||_p\leq\rho} \Big[ L(\theta+v; D_{tr}) \Big] - L(\theta; D_{tr}). 
\end{split} 
\end{align}
In this work, we focus on the definition of sharpness as in Eq.\ref{eq:sharpness}. Note that there are other sharpness definitions~\citep{andriushchenko2022understanding}, we will investigate them as one future work. Given this definition of sharpness, previous works~\citep{foret2020sharpness, wen2023does} build the theoretical and empirical connections between sharpness and model generalization. Generally, a smaller sharpness indicates a better generalization performance.

\vspace{-0.2cm}
\subsection{Data Poisoning Attacks}\label{sec:definition2}
\vspace{-0.1cm}
In this subsection, we give the definition of poisoning attacks we studied. Given a training set $D_{tr}$ with $n$ samples, we assume that the attacker selects a subset $D_q$ from $D_{tr}$ which takes an $\epsilon \in [0,1]$  percentage of $D_{tr}$, and replaces it with a poisoning set $D_p$. Usually, the samples in $D_p$ are from modifying samples in $D_q$. During the re-training stage, the model is trained from scratch on the perturbed training set $(D_{tr} - D_q) \cup D_p$ and we denote it as ``$D_{tr} + D_p$'' for simplicity. Although different attacks may have different purposes and formulations, we give a general formulation as in Eq.\ref{eq: general poison}: the attacker aims to find the poisoning samples $D_p$, such that the trained model has an optimized malicious behavior towards data samples from a victim set $D_T$:
\begin{equation}\label{eq: general poison}\small
\min_{D_p} ~~ Q(\theta^*, D_T),~~\text{s.t.}~~ \theta^* = \argmin_{\theta} L(\theta; D_{tr}+D_p)  \end{equation}
where $\mathcal{C}$ denotes the constraint of the poisoning samples. Notably, the design of ``\textit{\textbf{adversarial objective}}'' $Q(\cdot)$ and ``\textit{\textbf{victim set}}'' $D_T$ is based on the purpose of the adversary. By giving different formulations of $Q(\cdot)$ and definitions of $D_T$, the attacker can achieve various adversarial goals. For example, in targeted attacks~\citep{shafahi2018poison, geiping2020witches}, they aim to cause the model to misclassify a certain test sample (or set of test samples). Thus, they define the victim set $ D_T = \{(x_{i}^{vic},y^{vic})\}_{i=1}^k$ and the adversarial objective as:
\begin{align}\label{eq:ta}\small
    Q_t(\theta^*,  D_T) = \sum_{(x_i,y_i) \in D_T} l(f(x_i; \theta^*), y^{obj}),
\end{align} 
where $y^{obj}$ is a designated class different from $y^{vic}$. In Appendix~\ref{app:obj}, we provide a detailed introduction to the threat models of two other types of poisoning attacks, such as backdoor attacks~\citep{wu2022backdoorbench} and un-targeted attacks~\citep{huang2021unlearnable}.
\vspace{-0.3cm}
\section{Method}\label{sec:method}
\vspace{-0.2cm}

In this section, we introduce our proposed method Sharpness-aware Data Poisoning Attack (\ourmodel). In Section~\ref{sec:our_method}, we will first introduce \ourmodel in general form. In Section~\ref{sec:our_method1}, we introduce how to achieve \ourmodel in targeted and backdoor attacks. In Section~\ref{sec:our_method2}, we introduce \ourmodel in untargeted attacks.
From our discussion, we show the idea of leveraging sharpness is general and lightweight to be incorporated into various types of poisoning attacks. 

\vspace{-0.2cm}
\subsection{The Objective of 
\ourmodel}\label{sec:our_method}
\vspace{-0.1cm}


As discussed in  Section~\ref{sec:intro} \& \ref{sec:related}, one major challenge for poisoning attacks originates from the existence of multiple solutions during re-training. Therefore, we first define the \textit{\textbf{Poisoned Model Space}}, denotes as $\Theta_p$, which is the set of all models that are trained on a poisoned dataset $D_{tr}+D_p$ and have a small training loss: 
\begin{equation}
    \Theta_p= \{\theta: L(\theta; D_{tr}+D_p) \leq \tau \}
\end{equation}
In practice, the space $\Theta_p$ composes the models that are trained from different training algorithms, initializations and architectures. In the space $\Theta_p$, we 
desire to find the \textit{\textbf{Worst-case Poisoned Model}}, which is the model that has the worst poisoning effect $\theta' = \argmax_{\theta\in \Theta_p}~Q(\theta; D_{T})$. Ideally, we aim to optimize the poisoning effect of $\theta'$ to overcome the re-training uncertainty:
\begin{align}\label{eq:worst poison}\small
\min_{D_p } ~& Q(\theta'; D_{T}),\\ 
\text{where~} \theta' = \argmax~& Q(\theta; D_{T}), \text{~s.t.~}  L(\theta; D_{tr}+D_p) \leq \tau.
\end{align}
By solving this problem, we can find poisoning samples $D_p$ such that the worst poisoned model $\theta'$ can have an optimized poisoning effect (with a small $Q(\theta'; D_T)$). Therefore, for the models that are re-trained on the poisoned dataset, the poisoning effect is likely to persist, and our experiments in Section \ref{sec:exp} provide empirical evidence. Admittedly, it is also hard to solve the above problem exactly. As a solution, we propose to approximate the term $Q(\theta';D_T)$, by leveraging the existing notion of model sharpness~\citep{foret2020sharpness}. In detail, given a fixed $D_p$, we approximate $Q(\theta'; D_T)$ by:
\begin{align}\label{eq:sam_attack1}
Q(\theta'; D_T) \approx \max_{\|v\|_p\leq\rho} Q(\theta^*+v; D_T),~\text{where}~ \theta^* = \argmin_{\theta\in\Theta} L(\theta; D_{tr}+D_p)
\end{align}
In details, we first trained the model $\theta^*$ on $(D_{tr} + D_p)$ (i.e., via ERM), so that $\theta^*$ satisfies the constraint $L(\theta^*; D_{tr} + D_p) \leq \tau$. Then, we locally maximize the term  $Q(\theta^*+v)$ by perturbing the model parameter $\theta^*$ with a vector $v$ (which has a limited norm $\|v\|_p\leq\rho$). In this way, the perturbed model $\theta^* + v$  has a worse poisoning effect compared to $\theta^*$, but it is still likely to fall into the space $\Theta_p$ (because $\theta^* + v$ is not far away from $\theta^*$). 
Remarkably, the form of the objective Eq.~\ref{eq:sam_attack1} resembles the definition of model sharpness in Eq.~\ref{eq:sharpness}, while we focus on adversarial loss which is distinct from the sharpness on the standard loss landscape.
Therefore, we call the term in Eq.~\ref{eq:sam_attack1} as \textit{\textbf{Sharpness-aware adversarial objective}}, and we desire to find $D_p$ to optimize this objective:
\begin{align}\label{eq:sharp_targeted}
\min_{D_p}~ \max_{\|v\|_p\leq \rho}~ Q(\theta^*+v; D_T), ~\text{s.t.}~~ \theta^* = \argmin_{\theta \in \Theta} L(\theta; D_{tr}+D_p)
\end{align} 
In general, the formulation of \ourmodel can be seen as a modification of existing attacks: it replaces the adversarial objective (Eq.~\ref{eq: general poison}) of existing attacks with {{Sharpness-aware adversarial objective}}. Intuitively, compared with the traditional adversarial objective (Eq.~\ref{eq: general poison}), in \ourmodel, the perturbation on the inner minima $\theta^*$ with $v$ enables the algorithm to escape from the poisoned models where the poisoning effect is unstable. Thus, \ourmodel has the potential to induce the re-trained models to have a stable poisoning effect under the uncertainty of the re-training process. Notably, in our proposed approximation in Eq.\ref{eq:sam_attack1}, we locally search for a  worst model with the same model architecture of a trained model $\theta^*$, which does not explicitly consider the uncertainty of different architectures. However, empirical results in Section~\ref{sec:exp3} demonstrate that this strategy can also help improve the poisoning effect under if there is model architecture shift.

\vspace{-0.2cm}
\subsection{\ourmodel in Targeted /Backdoor Attacks}\label{sec:our_method1}
\vspace{-0.1cm}

In this subsection, we take one representative algorithm~\citep{geiping2020witches} in targeted attacks as an example, to
show \ourmodel can be incorporated into existing attacks without requiring additional attacker's knowledge or computational overhead. In the work Gradient Matching~\citep{geiping2020witches}, they assume the attacker is targeting on a victim set $D_T$, and the attacker has other information including a
pre-trained model $\theta^*$, which is trained on (or part of) the clean training set $D_{tr}$ and a loss function $L(\cdot)$. In their attack, they solve the following problem to find the set $D_p$ satisfying:
\begin{equation}\label{eq:gm}\small
\begin{aligned}
\argmin_{D_p}  \Big(1-\frac{\langle\nabla_{\theta}Q(\theta^*;D_T),\nabla_{\theta}L(\theta^*;D_{tr}+D_p)\rangle}{\|\nabla_{\theta}Q(\theta^*;D_T)\|\cdot \|\nabla_{\theta}L(\theta^*;D_{tr}+D_p)\|} \Big).
\end{aligned}
\end{equation}
It finds a poisoning set $D_p$, such that the gradient of the training loss, $\nabla_\theta L(\theta^*; D_{tr} + D_p)$, has a maximized alignment with $\nabla_\theta Q(\theta^*;D_T)$. In this way, during the re-training process, the model is updated by $\nabla_\theta L(\theta^*; D_{tr} + D_p)$ and likely to minimize $Q(\theta^*; D_T)$ and achieve the attack goal successfully. In our method \ourmodel, if we denote the sharpness-aware poisoning objective as $Q^S(\theta^*; D_T) = \max_{\|v\|_p\leq \rho}~ Q(\theta^*+v; D_T)$ for simplicity, the objective of \ourmodel is to solve:
\begin{equation}\label{eq:gm sharp targeted}\small
\begin{aligned}
\argmin_{D_p}  \Big(1-\frac{\langle\nabla_{\theta}Q^S(\theta^*;D_T),\nabla_{\theta}L(\theta^*;D_{tr}+D_p)\rangle}{\|\nabla_{\theta}Q^S(\theta^*;D_T)\|\cdot \|\nabla_{\theta}L(\theta^*;D_{tr}+D_p)\|} \Big),
\end{aligned}
\end{equation}
which directly replaces the adversarial objective in Eq.\ref{eq:gm} to be sharpness aware adversarial objective in Eq.\ref{eq:gm sharp targeted}. 
In our algorithm to calculate the term $\nabla_\theta Q^S(\theta^*; D_T)$, we follow the approach in the previous work~\citep{foret2020sharpness} to first approximate $Q^S(\theta^*; D_T)$ by leveraging a first-order method for $\hat{v}$:
\begin{align}\label{eq:estimate_v}
\hat{v}=\rho \cdot \text{sign}\Big(\nabla_{\theta}Q(\theta^*;D_T)\Big)|\nabla_{\theta}Q(\theta^*;D_T)|^{q-1}\left(\|\nabla_{\theta}Q(\theta^*;D_T)\|^q_q\right)^{1/p},
\end{align}
where $1/p + 1/q = 1$ and we consider $p=2$ as illustrated in \citep{foret2020sharpness} if not specified. 
Then, we can have the approximation to calculate $\nabla_\theta Q^S(\theta^*;D_T)$ via replacing $\theta^*$ with $\theta^*+\hat{v}$:
\begin{equation}\label{eq:gradient est}
    \nabla_{\theta}Q^S(\theta^*;D_T)\approx \nabla_{\theta}Q(\theta;D_T)|_{\theta = \theta^*+\hat{v}}
\end{equation}
In this way, by fixing $\nabla_{\theta}Q^S(\theta^*;D_T)$, we can solve Eq.\ref{eq:gm sharp targeted} to find the poisoning samples $D_p$ via gradient descent. Notably, the computation in Eq.\ref{eq:estimate_v}\&\ref{eq:gradient est} does not require additional attacker's knowledge compared to original Gradient Matching. In Appendix~\ref{app: target}, we provide the detailed algorithm of \ourmodel in Gradient Matching. We also analyze its computational efficiency to show \ourmodel does not introduce great computational overhead. Besides, the similar strategy of \ourmodel and algorithm can also be applied in backdoor attacks, given the similarity between the objectives and viable solutions of targeted attacks and backdoor attacks (see more discussions in Appendix~\ref{app: backdoor}).


\vspace{-0.3cm}
\subsection{\ourmodel in Un-targeted Attacks}\label{sec:our_method2}
\vspace{-0.2cm}
For untargeted attacks, similar strategy to replace the (investigated) model with a worse model during generating the poisoning samples also helps.
Note that conducting un-targeted attacks for DNN models with a small poisoning budget $\epsilon$ is usually considered to be a hard problem~\citep{munoz2017towards}. Existing feasible solutions~\citep{fowl2021adversarial, ren2022transferable} are motivated from a ``data protection perspective'' (which are often referred as ``un-learnable examples''). In detail, they perturb the whole (or at least a large portion of) training set, such that the trained classifiers cannot effectively learn the true data distribution and
have a low accuracy on the true data. Take the method \textbf{Error-Min} ~\citep{huang2021unlearnable} as an example, to induce the model to have a low accuracy, it generates ``shortcut'' perturbations
$\delta_i$ for each training sample to solve the following bilevel problem, such that the model trained on this perturbed dataset has a minimized loss:
\begin{align}\label{eq:emax}\small
  \min_{\theta\in \Theta} \min_{\{\delta_i\}_{i=1, ...,n}}  ~\Big[ \sum_{(x_i,y_i)\in D_{tr}} l\big(f(x_i + \delta_i;\theta), y_i\big)\Big],
\end{align} 
As a consequence, the found $\delta_i$ via solving Eq.~\ref{eq:emax} has patterns with a strong correlation to the labels $y_i$. The models trained on the perturbed dataset can predict $(x_i +\delta_i)$ to be $y_i$ mostly based on the information of $\delta_i$, and prohibit the model to learn useful knowledge from the clean samples in ${D}_{tr}$. In the original algorithm proposed by~\citep{huang2021unlearnable}, the problem in Eq.\ref{eq:emax} is solved by updating the model parameter $\theta$ and training set perturbation $\{\delta_i\}_{i=1}^n$ alternatively. See Algorithm~\ref{algo:aa1} (without \blue{SAPA}), they first update the model parameter $\theta$ with $M$ steps (step 1) and then update data perturbation $\delta_i$ for $T$ steps (step 3). For our method, similar to the high-level idea as in targeted attacks, \ourmodel introduces one more step (step 2) to replace the model $\theta$ with $\theta + \hat{v}$ during generating the poisoning samples. Notably, a similar strategy can also be adapted to other un-targeted attacks, i.e., Error-Max~\cite{fowl2021adversarial}. We provide its detailed algorithm in Appendix~\ref{app:algo}.

\vspace{-0.2cm}

\RestyleAlgo{ruled}
\begin{algorithm}[h]
\caption{Error-min\blue{+SAPA}}
\label{algo:aa1}
\SetKwInOut{Input}{Input}
\SetKwInOut{Output}{Output}
\Input{Clean training set $\{(x_i,y_i)\}_{i=1}^n$; Optimization step $T$ and $M$; epochs $E$.}
\Output{Sample-wise perturbation $D_p=\{\delta_i\}_{i=1}^n$}
Randomly initialize perturbation $D_P = \{\delta_i\}_{i=1}^{n}$\\
\For{epoch in $1,...,E$}
{
1. \For{$m$ in $1,...,M$}{ Update $\theta$ via minimizing $\sum_{(x_i,y_i)\in D_{tr}} l\big(f(x_i + \delta_i;\theta), y_i\big)$}
2. \blue{\textit{SAPA:}} Fix $\theta$, $D_P$, find worst direction $\hat{v}$ to maximize {\small $\sum_{(x_i,y_i)\in D_{tr}} l(f(x_i+\delta_i;\theta+\hat{v}), y_i)$} \\
3. \For{$t=1,...,T$ steps}{Update $\delta_i$ via gradient descent on the fixed model $f(\cdot;\theta+\hat{v})$
}
}
\end{algorithm}

\vspace{-0.3cm}
\section{Experiment}\label{sec:exp}
\vspace{-0.4cm}


In this section, we conduct experiments to show that \ourmodel can improve existing methods when it is applied to targeted, backdoor and un-targeted attacks in Section~\ref{exp:targeted}, \ref{exp:backdoor} and \ref{exp:untargeted}, respectively. For all methods in experiments, we obtain the poisoning samples via ResNet18 models. We report the results when the re-training is under ResNet18 and ResNet50, to have a thorough comparison with previous works (many baseline methods are also evaluated in the same setting).  
In Section.\ref{sec:exp3} we evaluate the robustness of \ourmodel under various re-training settings including different training algorithms (including some defense strategies) and several more model architectures. In Section~\ref{sec:exp2}, we compare \ourmodel with baselines incorporating \textit{ensemble} and \textit{re-initialization} technique to illustrate the superiority of \ourmodel in both effectiveness and efficiency. 
Through this section, we focus on image classification tasks on benchmark datasets CIFAR10 and CIFAR100.  Meanwhile, we provide additional empirical results of the dataset SVHN in Appendix~\ref{app:exp}. 
For our method \ourmodel, we set the radius $\rho$ (Eq.~\ref{eq:sharp_targeted} in Section~\ref{sec:method}) to be $0.05$ in all experiments. We provide detailed implementation in Appendix~\ref{app:algo} for all \ourmodel-based algorithms\footnote{Code is available in \href{}{https://github.com/PengfeiHePower/SAPA}}.

\vspace{-0.2cm}
\subsection{Performance of \ourmodel in Targeted Attacks}\label{exp:targeted}
\vspace{-0.3cm}

\textbf{Experiment Setup}. In this experiment, we focus on targeted attack goal to cause the poisoned model to misclassify one victim sample. During attacks, under each dataset, we assume that the attacker randomly chooses a small proportion of training dataset with ``\textit{poisoning ratio}'' $\epsilon =1\%,0.2\%$, and inserts unnoticeable perturbations (whose $l_\infty$ norm is limited by ``\textit{perturbation budget}'' $16/255,8/255, 4/255$) on each of them. After crafting the poisoned dataset, the model is randomly initialized and re-trained from scratch, via SGD for 160 epochs with an initial learning rate of 0.1 and decay by 0.1 at epochs 80 and 120. For each setting, we repeat the experiment 50 times and report the average performance. Multiple-target attacks are also considered in Appendix~\ref{app:exp}.
More details of the implementation can be found in Appendix.\ref{app: target}.

\vspace{-0.1cm}
\textbf{Baselines}. We compare \ourmodel with representative baselines, including Bullseye\citep{aghakhani2021bullseye}, Poison-Frog\citep{shafahi2018poison}, Meta-Poison~\citep{huang2020metapoison}, and  Grad-Match~\citep{geiping2020witches}. Notably, our method is incorporated to the method Grad-Match~\citep{geiping2020witches}. For other baselines, MetaPoison leverages meta learning~\citep{vilalta2002perspective} to unroll the training pipeline and ensembles multiple models. Poison-frog and Bullseye, also known as ``feature collision'', generate poisoning samples with representations similar to those of the victim. In Appendix~\ref{app: review}, we provide detailed discussions of these baselines.

\vspace{-0.1cm}
\textbf{Performance comparison}. In Table~ \ref{tab:result_targeted_1} we report the ``Success Rate'' which is the probability that the targeted sample is successfully classified to be the designated wrong label $y^{obj}$. From the results, we can see that \ourmodel consistently outperforms baselines. The advantage of \ourmodel is obvious, especially when the attacker's capacity is limited (i.e., the perturbation budget is $8/255$ or $4/255$). Additional results on ResNet50 can be found in Appendix~\ref{app:exp}.

\begin{table*}[h!]
\centering
\vspace{-0.2cm}
\caption{\small Success Rate in Targeted Attacks(with standard error reported).}
\vspace{-0.2cm}
\resizebox{0.85\textwidth}{!}
{
\begin{tabular}{c |c| cc | cc | cc } 
\hline
\hline
  & &   \multicolumn{2}{c|}{$16/255$}   &   \multicolumn{2}{c|}{$8/255$}   &   \multicolumn{2}{c}{$4/255$} \\
& & $\epsilon=1\%$ & 
$\epsilon=0.2\%$  & $\epsilon=1\%$ &$\epsilon=0.2\%$ & $\epsilon=1\%$ &$\epsilon=0.2\%$ \\
\hline
\multirow{4}{*}{\textbf{CIFAR10}} &\textbf{Bullseye} &3.7$\pm$1.6&1.1$\pm$0.4&1$\pm$0.3&0$\pm$0&0$\pm$0&0$\pm$0\\
&\textbf{Poison-Frog}& 1.3$\pm$0.8&0$\pm$0&0$\pm$0&0$\pm$0&0$\pm$0&0$\pm$0 \\
&\textbf{Meta-Poison}&42.5$\pm$2.8&30.7$\pm$2.3&28.1$\pm$2.5&19.4$\pm$2.1&5.2$\pm$1.7&3.9$\pm$1.2 \\
&\textbf{Grad-Match}&72.9$\pm$3.4 \footnotemark &63.4$\pm$3.8&35.4$\pm$3.1&26.8$\pm$3.4&10.3$\pm$2.6&6.4$\pm$2.8\\
&{\textbf{SAPA+Grad-Match}}&\blue{80.1$\pm$3.5}&\blue{70.8$\pm$2.7}&\blue{48.4$\pm$2.8}&\blue{31.5$\pm$3.1}&\blue{16.7$\pm$2.4}&\blue{11.2$\pm$1.5}\\
\hline
\multirow{4}{*}{\textbf{CIFAR100}} &\textbf{Bullseye}& 2.1$\pm$0.3&0$\pm$0&2.6$\pm$0.7&1.2$\pm$0.2&0.5$\pm$0.1&0.1$\pm$0.1\\
&\textbf{Poison-Frog}&1.0$\pm$0.8&0$\pm$0&1.1$\pm$0.4&0$\pm$0&0.2$\pm$0.1&0$\pm$0\\
&\textbf{Meta-Poison}&50.3$\pm$2.6&24.1$\pm$2.5&43.2$\pm$2.8&22.7$\pm$1.7&4.5$\pm$1.2&3.1$\pm$0.8 \\
&\textbf{Grad-Match}&90.2$\pm$3.1&53.6$\pm$2.5&62.1$\pm$3.5&33.4$\pm$4.4&11.3$\pm$2.5&7.4$\pm$1.9\\
&{\textbf{SAPA+Grad-Match}}&\blue{91.6$\pm$1.5}&\blue{74.9$\pm$2.2}&\blue{86.8$\pm$3.9}&\blue{52.6$\pm$3.6}&\blue{31.6$\pm$2.4}&\blue{12.1$\pm$1.2}\\
\hline\hline
\end{tabular}
}
\label{tab:result_targeted_1}
\end{table*}

\footnotetext{Results for baseline Grad-Match is lower than reported in the original paper, because the original paper only trains 40 epochs. Detailed analysis on the impact of epochs can be found in Appendix~\ref{app:exp}}

\vspace{-0.3cm}
\subsection{Performance of \ourmodel in Backdoor Attacks}\label{exp:backdoor}
\vspace{-0.3cm}

\textbf{Experiment setup}. 
In this subsection, we study the effectiveness of \ourmodel in backdoor attacks. In particular, we focus on the ``hidden-trigger'' setting~\citep{saha2020hidden, souri2022sleeper} where the attackers can only add imperceptible perturbations to the clean training samples. For backdoor attacks, the adversarial goal is to cause the samples from a victim class $y^{vic}$ to be wrongly classified as a designated class $y^{obj}$ by inserting triggers. Besides, we follow the setting in~\citep{souri2022sleeper} that the attacker adds imperceptible perturbations to samples in class $y^{obj}$. Therefore, in our study, we randomly choose the two different classes $y^{vic}, y^{obj}$ for poisoning sample generation. In this evaluation, we constrain that the attacker can only perturb 1\% of the whole training set and the re-training process resembles our settings in Section~\ref{exp:targeted}. All experiments are repeated for 50 times and we report the average success rate. More implementation details are in  Appendix.\ref{app: backdoor}

\textbf{Baselines}. In the experiment,  our proposed method \ourmodel is incorporated to Sleeper Agent~\citep{souri2022sleeper} which also leverages Gradient Match~\citep{geiping2020witches} to achieve the adversarial goal in backdoor attacks. 
We also show the results of the method, Hidden-Trigger Backdoor~\citep{saha2020hidden}, which optimizes the poison over images with triggers to preserve information of triggers, and the Clean-Label Backdoor method~\citep{turner2019label} leverages calculating adversarial examples~\citep{goodfellow2014explaining} to train a backdoored model. 

\textbf{Performance comparison}. Our results are shown in Table~\ref{tab:backdoor}, where we report the ``Success Rate'', which is the ratio of samples (with triggers) in $y^{vic}$ that are classified as $y^{obj}$ by the poisoned model. 
From the result, our method outperforms all baselines under all settings. Specifically, Hidden-trigger~\citep{saha2020hidden} and Clean-label~\citep{turner2019label} suffer from low ineffectiveness, as they are either designed for transfer learning or require control over the training process. 
Compared to these methods, \ourmodel shows effectiveness for different model architectures and perturbation budgets. 
In comparison with Sleeper Agent~\citep{souri2022sleeper}, which is also based on Gradient Matching~\citep{geiping2020witches}, our method can also have a clear improvement. Especially, under the perturbation budget 8/255 and CIFAR100 dataset, \ourmodel can obviously outperform Sleeper Agent.

\begin{table*}[h!]
\centering
\vspace{-0.2cm}
\caption{\small Success Rate in Backdoor Attacks on CIFAR10 and CIFAR100}
\vspace{-0.2cm}
\resizebox{0.8\textwidth}{!}
{
\begin{tabular}{c|c | cc | cc | cc } 
\hline
\hline
  & &   \multicolumn{2}{c|}{ResNet18}   &   \multicolumn{2}{c|}{ResNet50}   &   \multicolumn{2}{c}{VGG11} \\
& &$16/255$ & $8/255$ &$16/255$ &$8/255$ & $16/255$ &$8/255$ \\
\hline
\multirow{4}{*}{\textbf{CIFAR10}}
&\textbf{Hidden-trigger}&3.5$\pm$1.2&1.3$\pm$0.4&3.2$\pm$0.9&1.3$\pm$0.3&5$\pm$1.4&1.8$\pm$0.7\\
&\textbf{Clean-label}&2.7$\pm$1.1 \footnotemark &0.9$\pm$0.7&2.6$\pm$0.7&0.9$\pm$0.2&4.7$\pm$1.5&1.1$\pm$0.9\\
&\textbf{Sleeper Agent}&90.9$\pm$2.2&31.5$\pm$4.2&94.1$\pm$2.7&21.2$\pm$4.3&85.8$\pm$3.4&26.7$\pm$3.9\\
&{\textbf{SAPA+Sleeper Agent}}&\blue{98.1$\pm$2.6}&\blue{49.3$\pm$3.7}&\blue{98.4$\pm$4.9}&\blue{33.2$\pm$3.1}&\blue{94.3$\pm$2.6}&\blue{35.5$\pm$2.8}\\
\hline
\multirow{4}{*}{\textbf{CIFAR100}}
&\textbf{Hidden-trigger}&2.1$\pm$0.9&1.3$\pm$0.4&1.7$\pm$0.3&0.8$\pm$0.1&3.4$\pm$1.2&1.2$\pm$0.6\\
&\textbf{Clean-label}&1.5$\pm$0.7&0.9$\pm$0.2&1.2$\pm$0.4&0.4$\pm$0.1&2.6$\pm$0.3&0.8$\pm$0.2\\
&\textbf{Sleeper Agent}&58.3$\pm$3.9&26.7$\pm$4.3&47.2$\pm$4.5&18.5$\pm$3.7&41.6$\pm$3.2&12.9$\pm$2.6\\
&{\textbf{SAPA+Sleeper Agent}}&\blue{72.4$\pm$3.6}&\blue{41.8$\pm$3.2}&\blue{63.9$\pm$3.5}&\blue{31.4$\pm$3.2}&\blue{67.7$\pm$2.7}&\blue{30.3$\pm$2.3}\\
\hline\hline
\end{tabular}
}
\vspace{-0.1cm}
\label{tab:backdoor}
\end{table*}

\footnotetext{Hidden trigger and Clean-label also have lower success rates than in their original papers. This is because they are originally proposed for fine-tuning settings (by fixing the early layers), while we focus on end-to-end setting in this work. Our results are consistent with results in work~\citep{souri2022sleeper}.}

\vspace{-0.3cm}
\subsection{Performance of \ourmodel in Un-targeted Attacks}\label{exp:untargeted}
\vspace{-0.2cm}
\textbf{Experiment Setup}. The goal of un-targeted attacks is to degrade the models' test accuracy. 
We follow the line of existing works~\citep{huang2021unlearnable, fowl2021adversarial, ren2022transferable} (which are also called ``un-learnable examples'') to perturb a large portion of training samples ($50\%, 80\%, 100\%$) in CIFAR10 and CIFAR100 datasets, in order to protect the data against being learned by DNNs. In our experiment, we limit the  perturbation budget to $8/255$. We first generate poisoning samples targeting on ResNet18. Then we re-train the victim models under ResNet18 and ResNet50 following existing works~\citep{huang2020metapoison, fowl2021adversarial}. The training procedure for each model also resembles the settings in Section~\ref{exp:targeted}, and we repeat experiments for each setting 5 times and report the average performance. More details of the implementation are in Appendix~\ref{app: untarget}.

\textbf{Baselines}. We compare \ourmodel with representative ``un-learnable'' methods, such as Error-Min  \citep{huang2020metapoison}, Error-Max~\citep{fowl2021adversarial}, Separable Perturbation~\citep{yu2022availability}, and Autoregressive Perturbation~\citep{sandoval2022autoregressive}. We also report the clean performance which refers to the accuracy of models without poisoning attack.
Notably, our proposed method \ourmodel can be incorporated into either Error-Min or Error-Max, so we denote our method as ``Error-Min+\ourmodel'' and ``Error-Max + \ourmodel'' respectively. We provide more details of the algorithm of ``Error-Max+\ourmodel'' in Appendix~\ref{app:algo}.

\textbf{Performance comparison.} In Table~\ref{tab:result_untarget}, 
we report the accuracy of the re-trained model on the clean test dataset of CIFAR10 and CIFAR100, so a lower value indicates better attack performance. From the result \ourmodel can improve the poisoning effect for both Error-Min and Error-Max under all settings. For example, when incorporating \ourmodel to Error-Min, Error-Min+\ourmodel has a clear advantage, by reducing the accuracy to around 10\% when the poisoning ratio is $100\%$ in the CIFAR10 dataset. 
In other settings, Error-Min+\ourmodel can also manage to achieve a 2-4\% accuracy reduction compared to Error-Min. Similarly, Error-Max+\ourmodel is also demonstrated to have a consistent improvement over Error-Max. In addition, incorporating our strategy can boost Error-Max and Error-Min to achieve comparable performance with Autoregressive Perturbation which specifically targets on the vulnerability of CNNs.

\vspace{-0.2cm}
\begin{table*}[h!]
\centering
\caption{\small Test Accuracy of Models Trained on Poisoned Datasets via Un-targeted Attacks.}
\vspace{-0.2cm}
\resizebox{0.94\textwidth}{!}
{
\hspace{-1cm}
\begin{tabular}{c| ccc | ccc | ccc | ccc } 
\hline
\hline
 &   \multicolumn{6}{c|}{\textbf{CIFAR10}} &   \multicolumn{6}{c}{\textbf{CIFAR100}}   \\
\hline
  &   \multicolumn{3}{c|}{\textbf{ResNet18}}   &   \multicolumn{3}{c|}{\textbf{ResNet50}}   &   \multicolumn{3}{c|}{\textbf{ResNet18}} 
 & \multicolumn{3}{c}{\textbf{ResNet50}}\\
&100\% & 80\% &50\% &100\% & 80\% &50\%  &100\% & 80\% &50\%   &100\% & 80\% &50\%  \\
\hline
\textbf{Clean$^*$}& 94.8&	94.8&	94.8&	95.0&	95.0&	95.0&74.8&	74.8&	74.8&	75.2&	75.2&	75.2\\
\textbf{Separable.} &13.5&	86.3&	92.9&	14.9 &88.1&	93.2& 9.1&	57.1&	66.2&	8.4&	60.8&	66.7\\
\textbf{Autoregressive.} &11.8	& \blue{82.3} &	\blue{89.8} &	\blue{10.1} & \blue{83.6} &	\blue{90.3} & 4.2 &	\blue{51.6} &	\blue{64.7} &	\blue{4.3} & \blue{53.5} &	\blue{66.1}\\
\hline
\textbf{Error-Max}&  11.9&	88.2&	92.2&	12.8 &90.1&	93.9&4.8&	57.3&	66.9&	5.6&	58.3&	68.1\\
\textbf{Error-Max+SAPA} & \blue{9.6} &	84.6	&90.1&	10.9&85.7&	{91.3} & \blue{4.1}&	55.1&	{64.8} &	{4.9} &	{56.8} &	{66.9} \\
\hline
\textbf{Error-Min}& 21.2&	87.1&	93.4&	18.9& 89.5&	94.5&11.2&	56.9&	67.7&	10.8&	60.5&	70.3\\
\textbf{Error-Min+SAPA}& 10.9&	 {83.7}	&  {90.0}	&  {10.3} &  {85.2} &	91.8&8.7&	 {53.1} &	65.3&	9.5&	57.9&	67.6\\
\hline\hline
\end{tabular}
}
\label{tab:result_untarget}
\end{table*}

\vspace{-0.4cm}
\subsection{Robustness of re-training variants}\label{sec:exp3}
\vspace{-0.2cm}

We provide ablation studies on the effectiveness of \ourmodel under various re-training settings including different re-training algorithms, re-training schedules and model architectures. We also study the robustness of \ourmodel against adversarial training \citep{madry2017towards}. Notably, we only focus on the CIFAR10 dataset and ResNet18 (except for the studies on various architectures). For targeted and backdoor attacks, we specify the perturbation budget to $16/255$ with poisoning ratio $\epsilon = 1\%$. For un-targeted attacks, we specify the perturbation budget to be $8/255$ with poisoning ratio $\epsilon = 100\%$. 

\textbf{Various Re-training Algorithms}.
There are studies~\citep{schwarzschild2021just, ren2022transferable} demonstrating that many poisoning attacks can lose efficacy when faced with training algorithms beyond  Empirical Risk Minimization (ERM).
Therefore, we provide additional experiments to test the performance of \ourmodel under various re-training algorithms.
We mainly consider the algorithms including Cut-Out~\citep{devries2017improved}, Mix-Up~\citep{zhang2017mixup}, and  Sharpness-aware Minimization (SAM)~\citep{foret2020sharpness}, which is proposed to minimize model sharpness to improve model generalization. Two optimization methods---SGD and ADAM are also included. In Table~\ref{table1}, 
we compare \ourmodel with its backbone attack as baselines for each type of poisoning attack, and we use the same evaluation metric as previous subsections. From this table, we can see that our method remains outperforming the baseline attacks. 
Notably, among these re-training algorithms, Mix-Up shows an outstanding ability to reduce the poisoning effect for all attack methods that we studied. It may be because Mix-Up is a data augmentation strategy which drastically manipulates the training data distribution, which can weaken the poisoning effect of the injected poisoning samples.

\begin{table*}[h!]
\centering
\caption{\footnotesize SAPA vs Strongest Baselines under Re-training Scheme Variation. }
\vspace{-0.2cm}
\label{table1}
\hspace{0.2cm}
\resizebox{0.8\textwidth}{!}
{
\centering
\begin{tabular}{c | cc | cc | cc } 
\hline
\hline
  \multirow{2}{*}&   \multicolumn{2}{c|}{Un-targeted$(\downarrow)$}   &   \multicolumn{2}{c|}{Targeted$(\uparrow)$}   &   \multicolumn{2}{c}{Backdoor$(\uparrow)$} \\
 &Error-min & SAPA &GradMatch &SAPA & SleeperAgent &SAPA \\
 \hline
\textbf{SGD}&21.2&10.9 \blue{(-10.3)} &73.1&80.1 \blue{(+6.8)} &91.8&97.1 \blue{(+5.3)} \\
\textbf{ADAM}&19.7&10.4 \blue{(-9.3)} &80.2&85.4 \blue{(+5.2)} &92.3&97.7 \blue{(+5.4)} \\
\textbf{Cut-Out}&22.6&11.2 \blue{(-11.4)} & 82.4&90.3 \blue{(+7.9)} &97.8 & 100.0 \blue{(+2.2)} \\
\textbf{Mix-Up}&40.8&36.7 \blue{(-4.1)} & 58.4&	65.5 \blue{(+7.1)} &69.7&76.5 \blue{(+6.8)} \\
\textbf{SAM}&28.9&11.3 \blue{(-17.6)} & 74.3&	80.7 \blue{(+6.4)} & 79.9&85.3 \blue{(+5.4)} \\
\hline
  \multirow{2}{*} &   \multicolumn{2}{c|}{Un-targeted$(\downarrow)$}   &   \multicolumn{2}{c|}{Targeted$(\uparrow)$}   &   \multicolumn{2}{c}{Backdoor$(\uparrow)$} \\
   &Error-min & SAPA &GradMatch &SAPA & SleeperAgent &SAPA \\
  \hline
  \textbf{ResNet18}&21.2&10.9 \blue{(-10.3)} &73.1&80.1 \blue{(+7.0)} &91.8&97.1 \blue{(+5.3)} \\  \textbf{MobileNetV2}&21.5&11.9 \blue{(-9.6)} &68.5&75.2 \blue{(+6.7)} &30.2&37.6 \blue{(+7.4)} \\
\textbf{VGG11}&35.3&20.2 \blue{(-15.1)} &42.9&47.6 \blue{(+4.7)} &31.9&37.4 \blue{(+5.5)} \\
\textbf{\textbf{ViT}}&40.6&36.5 \blue{(-4.1)} &36.2 & 41.5 \blue{(+5.3)} &24.7&26.3 \blue{(+1.6)} \\
\hline 
\hline
\end{tabular}
}
\end{table*}

\textbf{Various model architectures}.
We also test different model architectures after the generation of poisoned data for all three types of attacks. In detail, we test on MobileNetV2~\citep{sandler2018mobilenetv2}, VGG11~\citep{simonyan2014very} and pretrained Vision Transformer(ViT, \citep{dosovitskiy2020image}). Results in Table~\ref{table1} conclude that \ourmodel consistently improves the baseline methods, and shows a better ability to adapt to unknown model architectures during re-training. 

\begin{figure}[h]
\vspace{-0.3cm}
\hspace{-0.5cm}
\subfloat[\footnotesize Backdoor$(\uparrow)$ (Step)]{ 
\begin{minipage}[c]{0.25\textwidth}
\centering
\includegraphics[width = 1\textwidth]{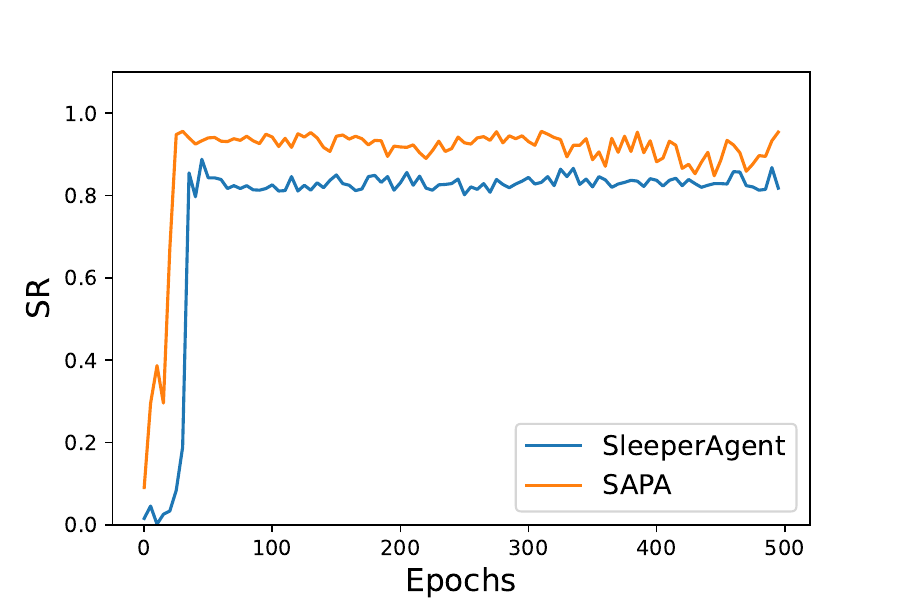}
\end{minipage}
}
\subfloat[\footnotesize Backdoor$(\uparrow)$ (Cyclic)]{ 
\begin{minipage}[c]{0.25\textwidth}
\centering
\includegraphics[width = 1\textwidth]{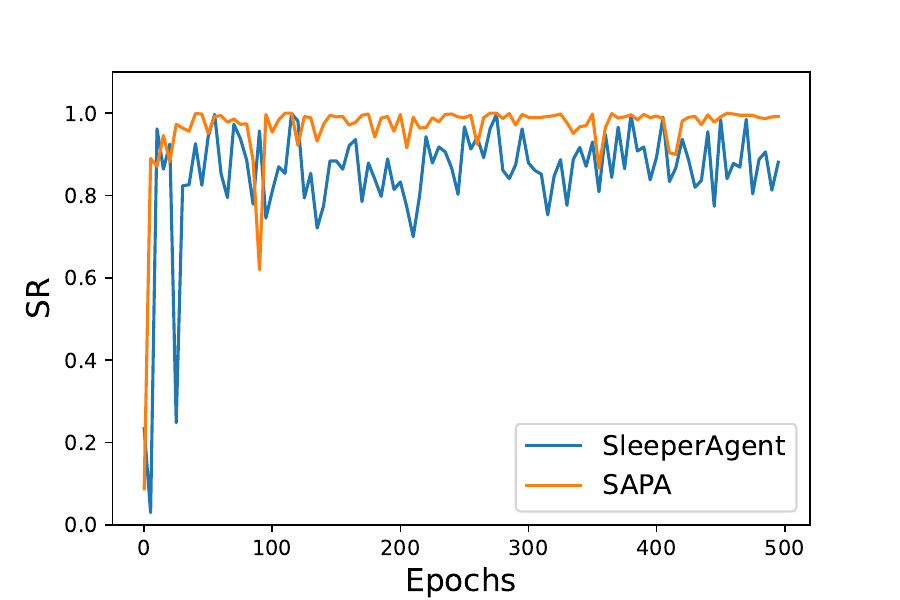}
\end{minipage}
}
\subfloat[\footnotesize Un-targeted$(\downarrow)$ (Step)]{ 
\begin{minipage}[c]{0.25\textwidth}
\centering
\includegraphics[width = 1\textwidth]{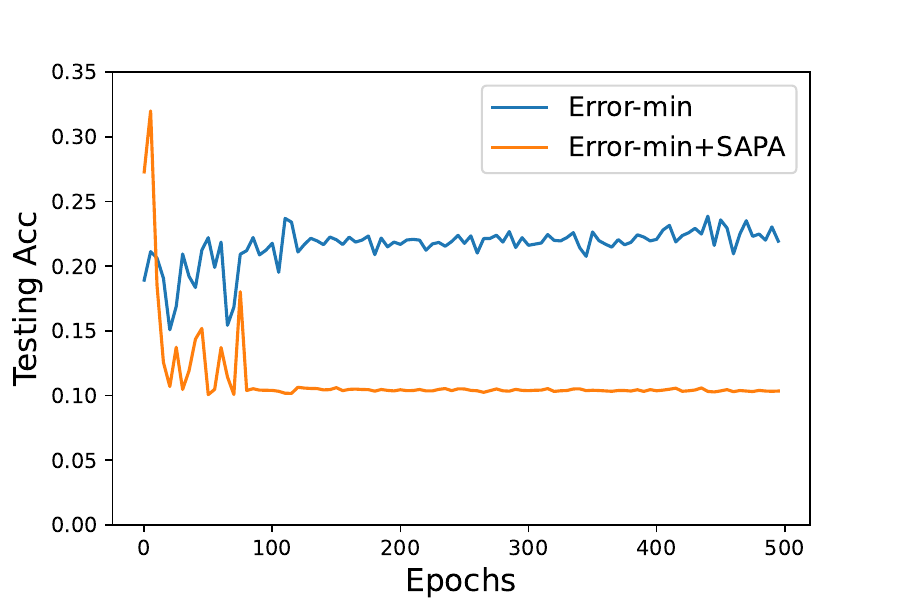}
\end{minipage}
}
\subfloat[\footnotesize Un-targeted$(\downarrow)$ (Cyclic)]{ 
\begin{minipage}[c]{0.25\textwidth}
\centering
\includegraphics[width = 1\textwidth]{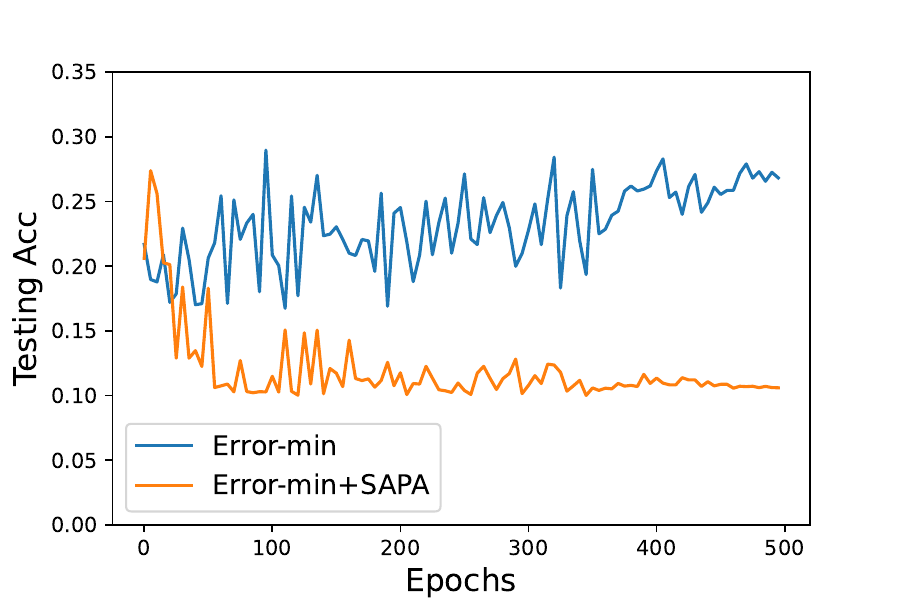}
\end{minipage}
}
\vspace{-0.2cm}
\caption{\small The Curve of Poisoning Effect for Various Re-training Schedules}
\label{fig:fair}
\end{figure}

\textbf{Various Re-Training Schedules}. 
There are also studies~\citep{schwarzschild2021just, huang2020metapoison} suggesting that a different re-training epoch number or schedule can significantly break the poisoning effect.
Therefore, we provide additional experiments when: (1) the model is trained for 500 epochs and the learning rate is updated by ``steps'' similar to previous studies, and (2) the re-training learning rate is updated ``cyclicly''. In Figure~\ref{fig:fair}, we plot the curve of the poisoning effect of \ourmodel and baselines in the backdoor and un-targeted attacks. Note that we exclude targeted attacks because the poisoning effect is discrete in one model. From these figures, we can find that \ourmodel can stably converge to the point with a strong poisoning effect that consistently outperforms baselines. 

\begin{wraptable}{r}{5cm}
\vspace{-0.4cm}
\caption{\footnotesize Adv. Train.}
\vspace{-0.3cm}
\label{tab:adv}
\centering
\resizebox{0.3\textwidth}{!}
{
\begin{tabular}{c|cc}
    \hline
    & ERM & Adv Train. \\
    \hline
    Separable& 11.8& 75.2\\
    Autoregressive&13.5& 78.7\\
    \hline
    Error-min & 21.2 & 82.0 \\
    Error-min + SAPA & 10.9 & 73.2\\
    \hline
    Error-max & 11.9 & 80.3\\
    Error-max + SAPA & 9.6 & 72.5\\
    \hline
    \end{tabular}}
\vspace{-0.3cm}
\end{wraptable}
\textbf{Adversarial Training}. In reality, these attacks can be faced to defense strategies. In this part, take untargeted attack as an example, we conduct a case study to evaluate the resistance of \ourmodel against Adversarial Training~\citep{madry2017towards},
which is a strong defense for un-targeted attacks~\citep{tao2021better}. We test on \ourmodel as well as all baselines in un-targeted attack. From the results, we note that adversarial training can significantly increase the test accuracy to defend against all attack methods.  However, \ourmodel can can still outperform other attacks. Notably, although Adv.~Train.~is usually considered as a strong defense to untargeted attacks, it can naturally reduce the accuracy on CIFAR10 from $95\%$ to $85\%$ without any poisoning attacks, based on extensive previous studies~\citep{tsipras2018robustness}.

\vspace{-0.3cm}
\subsection{Efficiency vs Effectiveness Trade-off}\label{sec:exp2}
\begin{figure}[h]
\vspace{-0.2cm}
\hspace{-0.5cm}
\subfloat[\footnotesize Targeted]{ 
\begin{minipage}[c]{0.5\textwidth}
\centering
\includegraphics[width = 0.95\textwidth]{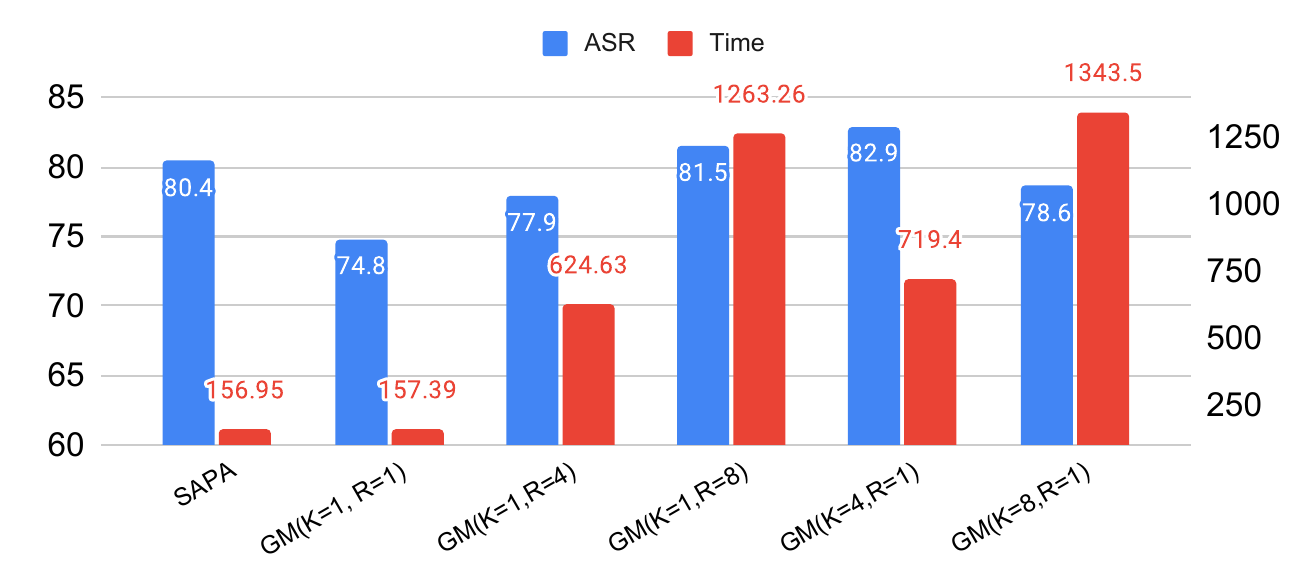}
\end{minipage}
\vspace{-0.2cm}
}
\subfloat[\footnotesize Backdoor]{ 
\begin{minipage}[c]{0.5\textwidth}
\centering
\includegraphics[width = 0.95\textwidth]{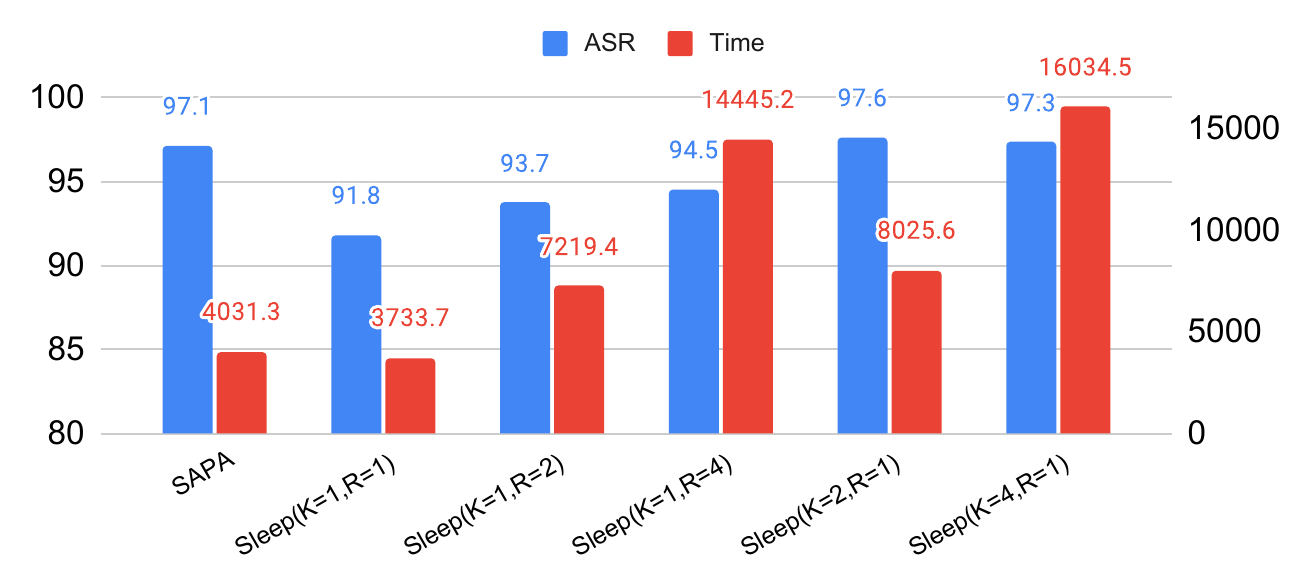}
\end{minipage}
\vspace{-0.2cm}
}
\vspace{-0.3cm}
\caption{\small ASR and running time for targeted attack (GM) and backdoor attack (Sleep.).}
\vspace{-0.3cm}
\label{fig:ER}
\end{figure}

\vspace{-0.3cm}
Previous works~\citep{huang2020metapoison} also leverage the
\textit{Ensemble} and \textit{Re-initialization} (E\&R) technique, to take various model architectures and initializations into consideration to handle re-training uncertainty. In this part, we compare the efficiency and effectiveness trade-off between \ourmodel and E\&R, when they are incorporated to existing attacks, such as Gradient Matching and Sleeper Agent. In Figure~\ref{fig:ER}, we report the attack successful rate (ASR) and computing time of \ourmodel and E\&R with different options ($K,R$ denote the number of ensembled model architectures and initializations respectively). Specifically, \ourmodel has higher ASR when $K$ and $R$ are small, and can still achieve comparable success rates when $K$ and $R$ are increasing. However, the running time grows dramatically for large $K$ and $R$ making them much less efficient than \ourmodel. This result shows \ourmodel demonstrate much better efficiency and effectiveness trade-off compared with E\&R. Notably, we exclude the result for untargeted attacks, as it generates poisoning samples for the whole training set, which makes E\&R extremely inefficient. 

\vspace{-0.2cm}
\section{Conclusion and Limitation}
\vspace{-0.3cm}

In this paper, we introduce a novel and versatile framework for data poisoning attacks that takes into account the model landscape sharpness. We apply this strategy to various types of poisoning attacks, including un-targeted, targeted, and backdoor attacks. Our experimental results demonstrate the superiority of our method compared to existing approaches. However, in this paper, our method focuses on image classification. Therefore, we will leave the relative studies, such as the poisoning attacks in self-supervised learning settings, and poisoning attacks in other domains such as texts and graphs for future investigation.

\section{Acknowledgement}
This research is supported by the National Science Foundation (NSF) under grant numbers CNS 2246050, IIS1845081, IIS2212032, IIS2212144, IOS2107215, DUE 2234015, DRL 2025244 and IOS2035472, the Army Research Office (ARO) under grant number W911NF-21-1-0198, the Home Depot, Cisco Systems Inc, Amazon Faculty Award, Johnson\&Johnson, JP Morgan Faculty Award and SNAP.

\bibliography{iclr2024_conference}

\begin{thebibliography}{62}
\providecommand{\natexlab}[1]{#1}
\providecommand{\url}[1]{\texttt{#1}}
\expandafter\ifx\csname urlstyle\endcsname\relax
  \providecommand{\doi}[1]{doi: #1}\else
  \providecommand{\doi}{doi: \begingroup \urlstyle{rm}\Url}\fi

\bibitem[Aghakhani et~al.(2021)Aghakhani, Meng, Wang, Kruegel, and Vigna]{aghakhani2021bullseye}
Hojjat Aghakhani, Dongyu Meng, Yu-Xiang Wang, Christopher Kruegel, and Giovanni Vigna.
\newblock Bullseye polytope: A scalable clean-label poisoning attack with improved transferability.
\newblock In \emph{2021 IEEE European Symposium on Security and Privacy (EuroS\&P)}, pp.\  159--178. IEEE, 2021.

\bibitem[Andriushchenko \& Flammarion(2022)Andriushchenko and Flammarion]{andriushchenko2022understanding}
Maksym Andriushchenko and Nicolas Flammarion.
\newblock Towards understanding sharpness-aware minimization, 2022.

\bibitem[Bard(2013)]{bard2013practical}
Jonathan~F Bard.
\newblock \emph{Practical bilevel optimization: algorithms and applications}, volume~30.
\newblock Springer Science \& Business Media, 2013.

\bibitem[Biggio et~al.(2012)Biggio, Nelson, and Laskov]{biggio2012poisoning}
Battista Biggio, Blaine Nelson, and Pavel Laskov.
\newblock Poisoning attacks against support vector machines.
\newblock \emph{arXiv preprint arXiv:1206.6389}, 2012.

\bibitem[Brown et~al.(2020)Brown, Mann, Ryder, Subbiah, Kaplan, Dhariwal, Neelakantan, Shyam, Sastry, Askell, et~al.]{brown2020language}
Tom Brown, Benjamin Mann, Nick Ryder, Melanie Subbiah, Jared~D Kaplan, Prafulla Dhariwal, Arvind Neelakantan, Pranav Shyam, Girish Sastry, Amanda Askell, et~al.
\newblock Language models are few-shot learners.
\newblock \emph{Advances in neural information processing systems}, 33:\penalty0 1877--1901, 2020.

\bibitem[Chaudhari et~al.(2019)Chaudhari, Choromanska, Soatto, LeCun, Baldassi, Borgs, Chayes, Sagun, and Zecchina]{chaudhari2019entropy}
Pratik Chaudhari, Anna Choromanska, Stefano Soatto, Yann LeCun, Carlo Baldassi, Christian Borgs, Jennifer Chayes, Levent Sagun, and Riccardo Zecchina.
\newblock Entropy-sgd: Biasing gradient descent into wide valleys.
\newblock \emph{Journal of Statistical Mechanics: Theory and Experiment}, 2019\penalty0 (12):\penalty0 124018, 2019.

\bibitem[Chen et~al.(2017)Chen, Liu, Li, Lu, and Song]{chen2017targeted}
Xinyun Chen, Chang Liu, Bo~Li, Kimberly Lu, and Dawn Song.
\newblock Targeted backdoor attacks on deep learning systems using data poisoning.
\newblock \emph{arXiv preprint arXiv:1712.05526}, 2017.

\bibitem[Colson et~al.(2007)Colson, Marcotte, and Savard]{colson2007overview}
Beno{\^\i}t Colson, Patrice Marcotte, and Gilles Savard.
\newblock An overview of bilevel optimization.
\newblock \emph{Annals of operations research}, 153:\penalty0 235--256, 2007.

\bibitem[DeVries \& Taylor(2017)DeVries and Taylor]{devries2017improved}
Terrance DeVries and Graham~W Taylor.
\newblock Improved regularization of convolutional neural networks with cutout.
\newblock \emph{arXiv preprint arXiv:1708.04552}, 2017.

\bibitem[Dinh et~al.(2017)Dinh, Pascanu, Bengio, and Bengio]{dinh2017sharp}
Laurent Dinh, Razvan Pascanu, Samy Bengio, and Yoshua Bengio.
\newblock Sharp minima can generalize for deep nets.
\newblock In \emph{International Conference on Machine Learning}, pp.\  1019--1028. PMLR, 2017.

\bibitem[Doan et~al.(2021)Doan, Lao, Zhao, and Li]{doan2021lira}
Khoa Doan, Yingjie Lao, Weijie Zhao, and Ping Li.
\newblock Lira: Learnable, imperceptible and robust backdoor attacks.
\newblock In \emph{Proceedings of the IEEE/CVF international conference on computer vision}, pp.\  11966--11976, 2021.

\bibitem[Dosovitskiy et~al.(2020)Dosovitskiy, Beyer, Kolesnikov, Weissenborn, Zhai, Unterthiner, Dehghani, Minderer, Heigold, Gelly, et~al.]{dosovitskiy2020image}
Alexey Dosovitskiy, Lucas Beyer, Alexander Kolesnikov, Dirk Weissenborn, Xiaohua Zhai, Thomas Unterthiner, Mostafa Dehghani, Matthias Minderer, Georg Heigold, Sylvain Gelly, et~al.
\newblock An image is worth 16x16 words: Transformers for image recognition at scale.
\newblock \emph{arXiv preprint arXiv:2010.11929}, 2020.

\bibitem[Dziugaite \& Roy(2017)Dziugaite and Roy]{dziugaite2017computing}
Gintare~Karolina Dziugaite and Daniel~M Roy.
\newblock Computing nonvacuous generalization bounds for deep (stochastic) neural networks with many more parameters than training data.
\newblock \emph{arXiv preprint arXiv:1703.11008}, 2017.

\bibitem[Foret et~al.(2020)Foret, Kleiner, Mobahi, and Neyshabur]{foret2020sharpness}
Pierre Foret, Ariel Kleiner, Hossein Mobahi, and Behnam Neyshabur.
\newblock Sharpness-aware minimization for efficiently improving generalization.
\newblock \emph{arXiv preprint arXiv:2010.01412}, 2020.

\bibitem[Fowl et~al.(2021)Fowl, Goldblum, Chiang, Geiping, Czaja, and Goldstein]{fowl2021adversarial}
Liam Fowl, Micah Goldblum, Ping-yeh Chiang, Jonas Geiping, Wojciech Czaja, and Tom Goldstein.
\newblock Adversarial examples make strong poisons.
\newblock \emph{Advances in Neural Information Processing Systems}, 34:\penalty0 30339--30351, 2021.

\bibitem[Franceschi et~al.(2018)Franceschi, Frasconi, Salzo, Grazzi, and Pontil]{franceschi2018bilevel}
Luca Franceschi, Paolo Frasconi, Saverio Salzo, Riccardo Grazzi, and Massimiliano Pontil.
\newblock Bilevel programming for hyperparameter optimization and meta-learning.
\newblock In \emph{International Conference on Machine Learning}, pp.\  1568--1577. PMLR, 2018.

\bibitem[Geiping et~al.(2020)Geiping, Fowl, Huang, Czaja, Taylor, Moeller, and Goldstein]{geiping2020witches}
Jonas Geiping, Liam Fowl, W~Ronny Huang, Wojciech Czaja, Gavin Taylor, Michael Moeller, and Tom Goldstein.
\newblock Witches' brew: Industrial scale data poisoning via gradient matching.
\newblock \emph{arXiv preprint arXiv:2009.02276}, 2020.

\bibitem[Goldblum et~al.(2022)Goldblum, Tsipras, Xie, Chen, Schwarzschild, Song, Madry, Li, and Goldstein]{goldblum2022dataset}
Micah Goldblum, Dimitris Tsipras, Chulin Xie, Xinyun Chen, Avi Schwarzschild, Dawn Song, Aleksander Madry, Bo~Li, and Tom Goldstein.
\newblock Dataset security for machine learning: Data poisoning, backdoor attacks, and defenses.
\newblock \emph{IEEE Transactions on Pattern Analysis and Machine Intelligence}, 45\penalty0 (2):\penalty0 1563--1580, 2022.

\bibitem[Goodfellow et~al.(2014)Goodfellow, Shlens, and Szegedy]{goodfellow2014explaining}
Ian~J Goodfellow, Jonathon Shlens, and Christian Szegedy.
\newblock Explaining and harnessing adversarial examples.
\newblock \emph{arXiv preprint arXiv:1412.6572}, 2014.

\bibitem[Gu et~al.(2019)Gu, Liu, Dolan-Gavitt, and Garg]{gu2019badnets}
Tianyu Gu, Kang Liu, Brendan Dolan-Gavitt, and Siddharth Garg.
\newblock Badnets: Evaluating backdooring attacks on deep neural networks.
\newblock \emph{IEEE Access}, 7:\penalty0 47230--47244, 2019.

\bibitem[Huang et~al.(2021)Huang, Ma, Erfani, Bailey, and Wang]{huang2021unlearnable}
Hanxun Huang, Xingjun Ma, Sarah~Monazam Erfani, James Bailey, and Yisen Wang.
\newblock Unlearnable examples: Making personal data unexploitable.
\newblock \emph{arXiv preprint arXiv:2101.04898}, 2021.

\bibitem[Huang et~al.(2020)Huang, Geiping, Fowl, Taylor, and Goldstein]{huang2020metapoison}
W~Ronny Huang, Jonas Geiping, Liam Fowl, Gavin Taylor, and Tom Goldstein.
\newblock Metapoison: Practical general-purpose clean-label data poisoning.
\newblock \emph{Advances in Neural Information Processing Systems}, 33:\penalty0 12080--12091, 2020.

\bibitem[Izmailov et~al.(2018)Izmailov, Podoprikhin, Garipov, Vetrov, and Wilson]{izmailov2018averaging}
Pavel Izmailov, Dmitrii Podoprikhin, Timur Garipov, Dmitry Vetrov, and Andrew~Gordon Wilson.
\newblock Averaging weights leads to wider optima and better generalization.
\newblock \emph{arXiv preprint arXiv:1803.05407}, 2018.

\bibitem[Keskar et~al.(2016)Keskar, Mudigere, Nocedal, Smelyanskiy, and Tang]{keskar2016large}
Nitish~Shirish Keskar, Dheevatsa Mudigere, Jorge Nocedal, Mikhail Smelyanskiy, and Ping Tak~Peter Tang.
\newblock On large-batch training for deep learning: Generalization gap and sharp minima.
\newblock \emph{arXiv preprint arXiv:1609.04836}, 2016.

\bibitem[Koh et~al.(2022)Koh, Steinhardt, and Liang]{koh2022stronger}
Pang~Wei Koh, Jacob Steinhardt, and Percy Liang.
\newblock Stronger data poisoning attacks break data sanitization defenses.
\newblock \emph{Machine Learning}, pp.\  1--47, 2022.

\bibitem[Li et~al.(2020{\natexlab{a}})Li, Gu, and Huang]{li2020improved}
Junyi Li, Bin Gu, and Heng Huang.
\newblock Improved bilevel model: Fast and optimal algorithm with theoretical guarantee.
\newblock \emph{arXiv preprint arXiv:2009.00690}, 2020{\natexlab{a}}.

\bibitem[Li et~al.(2020{\natexlab{b}})Li, Soltanolkotabi, and Oymak]{li2020gradient}
Mingchen Li, Mahdi Soltanolkotabi, and Samet Oymak.
\newblock Gradient descent with early stopping is provably robust to label noise for overparameterized neural networks.
\newblock In \emph{International conference on artificial intelligence and statistics}, pp.\  4313--4324. PMLR, 2020{\natexlab{b}}.

\bibitem[Liu et~al.(2020)Liu, Mu, Yuan, Zeng, and Zhang]{liu2020generic}
Risheng Liu, Pan Mu, Xiaoming Yuan, Shangzhi Zeng, and Jin Zhang.
\newblock A generic first-order algorithmic framework for bi-level programming beyond lower-level singleton.
\newblock In \emph{International Conference on Machine Learning}, pp.\  6305--6315. PMLR, 2020.

\bibitem[Liu et~al.(2021)Liu, Liu, Zeng, and Zhang]{liu2021towards}
Risheng Liu, Yaohua Liu, Shangzhi Zeng, and Jin Zhang.
\newblock Towards gradient-based bilevel optimization with non-convex followers and beyond.
\newblock \emph{Advances in Neural Information Processing Systems}, 34:\penalty0 8662--8675, 2021.

\bibitem[Lu et~al.(2022)Lu, Kamath, and Yu]{lu2022indiscriminate}
Yiwei Lu, Gautam Kamath, and Yaoliang Yu.
\newblock Indiscriminate data poisoning attacks on neural networks.
\newblock \emph{arXiv preprint arXiv:2204.09092}, 2022.

\bibitem[Madry et~al.(2017)Madry, Makelov, Schmidt, Tsipras, and Vladu]{madry2017towards}
Aleksander Madry, Aleksandar Makelov, Ludwig Schmidt, Dimitris Tsipras, and Adrian Vladu.
\newblock Towards deep learning models resistant to adversarial attacks.
\newblock \emph{arXiv preprint arXiv:1706.06083}, 2017.

\bibitem[Mei \& Zhu(2015)Mei and Zhu]{mei2015using}
Shike Mei and Xiaojin Zhu.
\newblock Using machine teaching to identify optimal training-set attacks on machine learners.
\newblock In \emph{Proceedings of the aaai conference on artificial intelligence}, volume~29, 2015.

\bibitem[Mu{\~n}oz-Gonz{\'a}lez et~al.(2017)Mu{\~n}oz-Gonz{\'a}lez, Biggio, Demontis, Paudice, Wongrassamee, Lupu, and Roli]{munoz2017towards}
Luis Mu{\~n}oz-Gonz{\'a}lez, Battista Biggio, Ambra Demontis, Andrea Paudice, Vasin Wongrassamee, Emil~C Lupu, and Fabio Roli.
\newblock Towards poisoning of deep learning algorithms with back-gradient optimization.
\newblock In \emph{Proceedings of the 10th ACM workshop on artificial intelligence and security}, pp.\  27--38, 2017.

\bibitem[Neyshabur et~al.(2017)Neyshabur, Bhojanapalli, and Srebro]{neyshabur2017pac}
Behnam Neyshabur, Srinadh Bhojanapalli, and Nathan Srebro.
\newblock A pac-bayesian approach to spectrally-normalized margin bounds for neural networks.
\newblock \emph{arXiv preprint arXiv:1707.09564}, 2017.

\bibitem[Nguyen \& Tran(2021)Nguyen and Tran]{nguyen2021wanet}
Anh Nguyen and Anh Tran.
\newblock Wanet--imperceptible warping-based backdoor attack.
\newblock \emph{arXiv preprint arXiv:2102.10369}, 2021.

\bibitem[Radford et~al.()Radford, Wu, Child, Luan, Amodei, Sutskever, et~al.]{radford2019language}
Alec Radford, Jeffrey Wu, Rewon Child, David Luan, Dario Amodei, Ilya Sutskever, et~al.
\newblock Language models are unsupervised multitask learners.

\bibitem[Radford et~al.(2021)Radford, Kim, Hallacy, Ramesh, Goh, Agarwal, Sastry, Askell, Mishkin, Clark, et~al.]{radford2021learning}
Alec Radford, Jong~Wook Kim, Chris Hallacy, Aditya Ramesh, Gabriel Goh, Sandhini Agarwal, Girish Sastry, Amanda Askell, Pamela Mishkin, Jack Clark, et~al.
\newblock Learning transferable visual models from natural language supervision.
\newblock In \emph{International conference on machine learning}, pp.\  8748--8763. PMLR, 2021.

\bibitem[Ramesh et~al.(2021)Ramesh, Pavlov, Goh, Gray, Voss, Radford, Chen, and Sutskever]{ramesh2021zero}
Aditya Ramesh, Mikhail Pavlov, Gabriel Goh, Scott Gray, Chelsea Voss, Alec Radford, Mark Chen, and Ilya Sutskever.
\newblock Zero-shot text-to-image generation.
\newblock In \emph{International Conference on Machine Learning}, pp.\  8821--8831. PMLR, 2021.

\bibitem[Ren et~al.(2022)Ren, Xu, Wan, Ma, Sun, and Tang]{ren2022transferable}
Jie Ren, Han Xu, Yuxuan Wan, Xingjun Ma, Lichao Sun, and Jiliang Tang.
\newblock Transferable unlearnable examples.
\newblock \emph{arXiv preprint arXiv:2210.10114}, 2022.

\bibitem[Saha et~al.(2020)Saha, Subramanya, and Pirsiavash]{saha2020hidden}
Aniruddha Saha, Akshayvarun Subramanya, and Hamed Pirsiavash.
\newblock Hidden trigger backdoor attacks.
\newblock In \emph{Proceedings of the AAAI conference on artificial intelligence}, volume~34, pp.\  11957--11965, 2020.

\bibitem[Sandler et~al.(2018)Sandler, Howard, Zhu, Zhmoginov, and Chen]{sandler2018mobilenetv2}
Mark Sandler, Andrew Howard, Menglong Zhu, Andrey Zhmoginov, and Liang-Chieh Chen.
\newblock Mobilenetv2: Inverted residuals and linear bottlenecks.
\newblock In \emph{Proceedings of the IEEE conference on computer vision and pattern recognition}, pp.\  4510--4520, 2018.

\bibitem[Sandoval-Segura et~al.(2022)Sandoval-Segura, Singla, Geiping, Goldblum, Goldstein, and Jacobs]{sandoval2022autoregressive}
Pedro Sandoval-Segura, Vasu Singla, Jonas Geiping, Micah Goldblum, Tom Goldstein, and David Jacobs.
\newblock Autoregressive perturbations for data poisoning.
\newblock \emph{Advances in Neural Information Processing Systems}, 35:\penalty0 27374--27386, 2022.

\bibitem[Schwarzschild et~al.(2021)Schwarzschild, Goldblum, Gupta, Dickerson, and Goldstein]{schwarzschild2021just}
Avi Schwarzschild, Micah Goldblum, Arjun Gupta, John~P Dickerson, and Tom Goldstein.
\newblock Just how toxic is data poisoning? a unified benchmark for backdoor and data poisoning attacks.
\newblock In \emph{International Conference on Machine Learning}, pp.\  9389--9398. PMLR, 2021.

\bibitem[Shaban et~al.(2019)Shaban, Cheng, Hatch, and Boots]{shaban2019truncated}
Amirreza Shaban, Ching-An Cheng, Nathan Hatch, and Byron Boots.
\newblock Truncated back-propagation for bilevel optimization.
\newblock In \emph{The 22nd International Conference on Artificial Intelligence and Statistics}, pp.\  1723--1732. PMLR, 2019.

\bibitem[Shafahi et~al.(2018)Shafahi, Huang, Najibi, Suciu, Studer, Dumitras, and Goldstein]{shafahi2018poison}
Ali Shafahi, W~Ronny Huang, Mahyar Najibi, Octavian Suciu, Christoph Studer, Tudor Dumitras, and Tom Goldstein.
\newblock Poison frogs! targeted clean-label poisoning attacks on neural networks.
\newblock \emph{Advances in neural information processing systems}, 31, 2018.

\bibitem[Simonyan \& Zisserman(2014)Simonyan and Zisserman]{simonyan2014very}
Karen Simonyan and Andrew Zisserman.
\newblock Very deep convolutional networks for large-scale image recognition.
\newblock \emph{arXiv preprint arXiv:1409.1556}, 2014.

\bibitem[Souri et~al.(2022)Souri, Fowl, Chellappa, Goldblum, and Goldstein]{souri2022sleeper}
Hossein Souri, Liam Fowl, Rama Chellappa, Micah Goldblum, and Tom Goldstein.
\newblock Sleeper agent: Scalable hidden trigger backdoors for neural networks trained from scratch.
\newblock \emph{Advances in Neural Information Processing Systems}, 35:\penalty0 19165--19178, 2022.

\bibitem[Sow et~al.(2022)Sow, Ji, Guan, and Liang]{sow2022primaldual}
Daouda Sow, Kaiyi Ji, Ziwei Guan, and Yingbin Liang.
\newblock A primal-dual approach to bilevel optimization with multiple inner minima, 2022.

\bibitem[Steinhardt et~al.(2017)Steinhardt, Koh, and Liang]{steinhardt2017certified}
Jacob Steinhardt, Pang Wei~W Koh, and Percy~S Liang.
\newblock Certified defenses for data poisoning attacks.
\newblock \emph{Advances in neural information processing systems}, 30, 2017.

\bibitem[Suya et~al.(2021)Suya, Mahloujifar, Suri, Evans, and Tian]{suya2021model}
Fnu Suya, Saeed Mahloujifar, Anshuman Suri, David Evans, and Yuan Tian.
\newblock Model-targeted poisoning attacks with provable convergence.
\newblock In \emph{International Conference on Machine Learning}, pp.\  10000--10010. PMLR, 2021.

\bibitem[Tao et~al.(2021)Tao, Feng, Yi, Huang, and Chen]{tao2021better}
Lue Tao, Lei Feng, Jinfeng Yi, Sheng-Jun Huang, and Songcan Chen.
\newblock Better safe than sorry: Preventing delusive adversaries with adversarial training.
\newblock \emph{Advances in Neural Information Processing Systems}, 34:\penalty0 16209--16225, 2021.

\bibitem[Tran et~al.(2018)Tran, Li, and Madry]{tran2018spectral}
Brandon Tran, Jerry Li, and Aleksander Madry.
\newblock Spectral signatures in backdoor attacks.
\newblock \emph{Advances in neural information processing systems}, 31, 2018.

\bibitem[Tsipras et~al.(2018)Tsipras, Santurkar, Engstrom, Turner, and Madry]{tsipras2018robustness}
Dimitris Tsipras, Shibani Santurkar, Logan Engstrom, Alexander Turner, and Aleksander Madry.
\newblock Robustness may be at odds with accuracy.
\newblock \emph{arXiv preprint arXiv:1805.12152}, 2018.

\bibitem[Turner et~al.(2019)Turner, Tsipras, and Madry]{turner2019label}
Alexander Turner, Dimitris Tsipras, and Aleksander Madry.
\newblock Label-consistent backdoor attacks.
\newblock \emph{arXiv preprint arXiv:1912.02771}, 2019.

\bibitem[Vilalta \& Drissi(2002)Vilalta and Drissi]{vilalta2002perspective}
Ricardo Vilalta and Youssef Drissi.
\newblock A perspective view and survey of meta-learning.
\newblock \emph{Artificial intelligence review}, 18:\penalty0 77--95, 2002.

\bibitem[Wen et~al.(2023)Wen, Ma, and Li]{wen2023does}
Kaiyue Wen, Tengyu Ma, and Zhiyuan Li.
\newblock How does sharpness-aware minimization minimize sharpness?, 2023.

\bibitem[Wu et~al.(2022)Wu, Chen, Zhang, Zhu, Wei, Yuan, and Shen]{wu2022backdoorbench}
Baoyuan Wu, Hongrui Chen, Mingda Zhang, Zihao Zhu, Shaokui Wei, Danni Yuan, and Chao Shen.
\newblock Backdoorbench: A comprehensive benchmark of backdoor learning.
\newblock \emph{Advances in Neural Information Processing Systems}, 35:\penalty0 10546--10559, 2022.

\bibitem[Xiao et~al.(2015)Xiao, Biggio, Brown, Fumera, Eckert, and Roli]{xiao2015feature}
Huang Xiao, Battista Biggio, Gavin Brown, Giorgio Fumera, Claudia Eckert, and Fabio Roli.
\newblock Is feature selection secure against training data poisoning?
\newblock In \emph{international conference on machine learning}, pp.\  1689--1698. PMLR, 2015.

\bibitem[Yu et~al.(2022)Yu, Zhang, Chen, Yin, and Liu]{yu2022availability}
Da~Yu, Huishuai Zhang, Wei Chen, Jian Yin, and Tie-Yan Liu.
\newblock Availability attacks create shortcuts.
\newblock In \emph{Proceedings of the 28th ACM SIGKDD Conference on Knowledge Discovery and Data Mining}, pp.\  2367--2376, 2022.

\bibitem[Zhang et~al.(2017)Zhang, Cisse, Dauphin, and Lopez-Paz]{zhang2017mixup}
Hongyi Zhang, Moustapha Cisse, Yann~N Dauphin, and David Lopez-Paz.
\newblock mixup: Beyond empirical risk minimization.
\newblock \emph{arXiv preprint arXiv:1710.09412}, 2017.

\bibitem[Zhao \& Lao(2022)Zhao and Lao]{zhao2022clpa}
Bingyin Zhao and Yingjie Lao.
\newblock Clpa: Clean-label poisoning availability attacks using generative adversarial nets.
\newblock In \emph{Proceedings of the AAAI Conference on Artificial Intelligence}, volume~36, pp.\  9162--9170, 2022.

\bibitem[Zhu et~al.(2019)Zhu, Huang, Li, Taylor, Studer, and Goldstein]{zhu2019transferable}
Chen Zhu, W~Ronny Huang, Hengduo Li, Gavin Taylor, Christoph Studer, and Tom Goldstein.
\newblock Transferable clean-label poisoning attacks on deep neural nets.
\newblock In \emph{International Conference on Machine Learning}, pp.\  7614--7623. PMLR, 2019.

\end{thebibliography}
\bibliographystyle{iclr2024_conference}

\appendix
\newpage
\appendix

\section{Boarder Impact}

Our research unveils a powerful attacking framework that has the potential to compromise security-critical systems in a stealthy manner. As machine learning models, particularly large models that require extensive training datasets, become increasingly prevalent and assume significant roles in various domains, the importance of clean training data cannot be overstated. It is imperative to prioritize the quality of data to ensure the success and reliability of these models.
By shedding light on the potential dangers associated with this threat model, our study aims to raise awareness about the importance of data security. We hope that our findings will serve as a catalyst for the development of stronger defenses against data poisoning attacks. Safeguarding against these threats requires a proactive approach and increased vigilance to protect the integrity and robustness of machine learning systems.

\section{Discussion of Existing Data Poisoning Attacks.}\label{app: review}
\vspace{-0.3cm}

{\subsection{Threat model and objectives}\label{app:obj}
In this section, we introduce the threat model, adversarial objective, and victim set for each type of data poisoning attack.

\textbf{Un-targeted Attacks}. In un-targeted attacks~\citep{steinhardt2017certified}, the attacker aims to cause the trained model with an overall low test accuracy. The attackers are assumed to have access to the training data and can perturb part~\citep{steinhardt2017certified} or the whole training data~\citep{fowl2021adversarial, huang2021unlearnable}. However, because the attacker usually does not have knowledge of test distribution and the training process of the victim model, most works~\citep{steinhardt2017certified, fowl2021adversarial, huang2021unlearnable} define the adversarial objective as the following to maximize the model error on the clean training set $D_{tr}$:
\begin{align}\label{eq:tua}
    Q_{ut}(\theta^*, D_{tr}) = - L(\theta^*; D_{tr}) 
\end{align}

\textbf{Targeted attacks}. In targeted attacks~\citep{shafahi2018poison}, the attacker aims to cause the trained model to misclassify a specified test sample or a subset of test samples. For example, they are targeting on a victim person and have knowledge of $k~ (k\geq 1)$ photographs of this person\footnotemark $\{(x_i^{vic},y^{vic})\}_{i=1}^k$. They aim to cause the model to misclassify the photos of this person while preserving the overall accuracy of the rest. The attackers are only allowed to perturb a small part of the training data and have no knowledge of the victim model including the initialization and training algorithm. Therefore, they define the victim set $ D_T = \{(x_{i}^{vic},y^{vic})\}_{i=1}^k$ and the adversarial objective as:
\begin{align}\label{eq:ta}
    Q_t(\theta^*,  D_T) = \sum_{(x_i,y_i) \in D_T} l(f(x_i; \theta^*), y^{obj}),
\end{align} 
where $y^{obj}$ is a designated class different from $y^{vic}$.

\footnotetext{We assume that the samples of the victim are from the same class $y^{vic}$, following most existing works~\citep{shafahi2018poison}.}


\textbf{Backdoor attacks}. In backdoor attacks~\citep{chen2017targeted, souri2022sleeper}, the attacker aims to take control of the model prediction by injecting samples with ``triggers''. In particular, if there is a trigger, such as a patch $p$, present in an image, the poisoned model will predict this sample to a specified class $y^{obj}$. Otherwise, the poisoned model will make a correct prediction. Similar to the targeted attack, attackers are only allowed to perturb a small part of training data and have no control of the training process of the victim model. In backdoor attacks, most works target on samples from a specific victim class $y = y^{vic}$ and define the victim set as $ D_{T} = \{(x,y)\in D_{tr} | y = y^{vic}\}$. During the attack, they aim to solve the adversarial objective: 
\begin{align}\label{eq:backdoor}\small
    Q_b(\theta^*,  D_T) = \sum_{(x_i ,y_i) \in D_T} l(f(x_i \oplus p; \theta^*), y^{obj})
\end{align}
where $x\oplus p$ denotes the process that $p$ is attached to a clean image $x$.
In this way, the poisoned model is likely to predict the samples with triggers to be class $y^{obj}$.}

\subsection{Algorithms}\label{app:b2}

In this section, we discuss the details of existing data poisoning attacks.

\textbf{Targeted attacks.} Targeted attacks insert poisons into the clean training aiming at misclassifying targets(samples or classes) as adversarial labels.  Fundamentally, targeted attacks can be formulated as a bi-level optimization problem in Eq.\ref{eq:ta}, which can be challenging especially for DNN because the inner problem has multiple minima. Some existing methods avoid solving it directly and apply heuristic approaches such as Feature Collision Attacks(i.e. Bullseye\citep{aghakhani2021bullseye}, poison-frog\citep{shafahi2018poison}) which manipulate the representation of victims to mislead models classify them as adversarial labels. These methods are well-suited for the transfer learning setting\citep{goldblum2022dataset}, where a model is pre-trained on clean data and fine-tuned on a smaller poisoned dataset. However, this brings a natural drawback in that its poisoning effect is only kept for one surrogate model(the pre-trained model), thus these methods can hardly work for the retrain-from-scratch scenario as the retrained model can be very different from the surrogate model. Another line of work tries to handle the bi-level optimization problem directly. For linear or convex models, the inner problem typically exhibits a single minimum. As a result, several methods such as those proposed by\citep{biggio2012poisoning, xiao2015feature} have been developed to successfully achieve the malicious objective. Theoretical analyses, such as the work by\citep{mei2015using}, have provided guarantees in these cases. However, these methods become ineffective for DNN. \citep{munoz2017towards} applies a method called “back-gradient descent” in which the inner problem is approximately solved using several steps of gradient descent, and the gradient-descent for the outer is conducted by back-propagating through the inner minimization routine. This method is both time and memory-consuming, thus impractical for models with multiple layers. MetaPoison\citep{huang2020metapoison} draws inspiration from \citep{franceschi2018bilevel, shaban2019truncated} and unrolling the training pipeline of inner. They also apply ``ensembling'' and ``network re-initialization'' to avoid overfitting to one single surrogate model and try to preserve the poisoning effect during re-training. However, the success of this unrolling  still requires single minima assumption\citep{liu2020generic}, leading to less effectiveness as shown in the empirical results. Grad-Match\citep{geiping2020witches} leverages a “gradient alignment” technique to solve the problem in Eq.\ref{eq:ta} and applies  ``poison re-initialization'' to select better poisons, but there still exists space for improvement as the formulation does not capture the nature of inner multiple minima well\citep{sow2022primaldual}. This is the reason why we rethink the formulation of targeted attacks and design a new objective in Eq.\ref{eq:worst poison}. 

\textbf{Backdoor attacks.} Backdoor attacks aim at inserting triggers into training samples and testing samples, such that triggered inputs will be misclassified. Note that there exist many types of backdoor attacks\citep{wu2022backdoorbench} with regard to factors of backdoor, such as attacker's capability(training controllable or not), characteristics of triggers(visibility, etc). In Eq.\ref{eq:backdoor}, we focus on so-called \textit{hidden-trigger clean-label} backdoor attacks, meaning attackers insert poisoned samples rather than triggers into the training data, and testing inputs attached with a trigger will be misclassified. Hidden-trigger\citep{saha2020hidden} generates poisoned images carrying information about triggered source images and confuses the model during fine-tuning. Clean-label\citep{turner2019label} leverages adversarial perturbations to cause a successful backdoored effect. However, both methods can hardly succeed in the retrain-from-scratch scenario, as Hidden-trigger is designed for transferring learning and Clean-label needs to control the retraining process. Sleeper Agent\citep{souri2022sleeper} incorporates the gradient matching technique to solve the problem in Eq.\ref{eq:backdoor}, and considers ``model restarts'' to adaptively update the model during the generation of poisons, but it can only cover a few models and limits its effectiveness. Our SAPA utilizes the sharpness objective to capture the nature of multiple-inner-minima, and is shown to better preserve the poisoning effect during retraining. {It is worth noting that there exists a line of work that consider a different scenario from end-to-end scenario discussed in this paper. Some of these attacks need additional assumptions that the attacker can control the re-training process such as WaNet~\citep{nguyen2021wanet} and LiRA~\citep{doan2021lira}; some conduct attacks without involving any models and require noticeable triggers, such as BadNet~\citep{gu2019badnets}. Therefore, this line of work is out of the main scope of our paper.}

\textbf{Un-targeted attacks.} Un-targeted attacks insert poisons into the clean training and the goal is to degrade the overall accuracy over the clean testing. Generally, it can be formulated as a bi-level optimization problem in Eq.\ref{eq:tua}. Early works\citep{biggio2012poisoning, steinhardt2017certified, koh2022stronger} primarily focused on simple models with desirable properties like linearity and convexity.  In these cases, the inner problem typically possesses a unique solution, and these methods demonstrate effective performance. Theoretical analysis\citep{steinhardt2017certified, suya2021model} have been carried out to establish the feasibility of untargeted attacks on such simple models. Nevertheless, the presence of non-convexity and multiple minima within the inner problem poses significant challenges for untargeted attacks on DNN models. \citep{munoz2017towards} applies “back-gradient descent”, which is expensive and infeasible for complex models. \citep{lu2022indiscriminate} borrows ideas from Stackelberg games and applies so-called ``total gradient descent ascent'' techniques to generate poison samples one by one. However, they can hardly preserve poisoning effects after retraining and have a subtle influence on clean testing. CLPA\citep{zhao2022clpa} takes advantage of 'naturally poisoned data' which refers to the wrongly predicted samples by the clean model and generates them using GAN. Nevertheless, this method can only be applied for transferring learning as it relies on the clean model and suffers from the multiple-inner-minima problem. MetaPoison\citep{huang2020metapoison} tries to handle the multiple-inner-minima problem using ensembling and re-initialization, but the performance is not very satisfactory. Error-max\citep{fowl2021adversarial} solves the problem in Eq.\ref{eq:tua} by leveraging adversarial examples, and Error-min\citep{huang2021unlearnable} inserts perturbations to build a relationship between labels and perturbations. These two methods only involve one surrogate model and can be improved through our proposed method. 
There also exists another line of work that generate perturbations without access to model or datasets, such as Synthetic Perturbations\citep{yu2022availability} and Autoregressive Perturbations\citep{sandoval2022autoregressive}. Though they do not solve the optimization problem directly and do not have the multiple-inner-minima problem, their application may be limited to specific settings(such as un-targeted attacks and CNN models), and not as general as ours.

\section{Implementation Details of All \ourmodel Methods}\label{app:algo}
In this subsection, we provide details of the implementation of \ourmodel methods.
\vspace{-0.25cm}
\subsection{Targeted attacks}\label{app: target}
\vspace{-0.25cm}
As introduced in Section.\ref{sec:our_method}, we solve the optimzation problem in Eq.\ref{eq:sharp_targeted} using Gradient Matching. To be more specific, we randomly choose one or a few target samples from a victim class(in testing) as the victim set $D_T$. At the same time, we randomly choose a percentage of samples $D_p$ from the training to be modified in the algorithm, and the percentage is referred to as \textit{poisoning ratio}. The attacking goal is to make the model trained on $D_{tr}+D_p$ misclassify samples in $D_T$ as a pre-specified adversarial class(different from the true class). To ensure the imperceptibility of poisons, we rigorously control the \textit{poisoning ratio} and \textit{perturbation budget}($L_{\infty}$ norm of the perturbation) and in our experiments, we consider the ratio of $1\%, 0.2\%$ and budget of $16/255,8/255, 4/255$. 

Given a pre-trained clean model, we first randomly initialize the perturbation for $D_p$, and during each optimization step, we compute the gradient of the sharpness-aware objective using Eq.\ref{eq:gradient est} and do one step gradient descent to minimize the similarity loss defined in Eq.\ref{eq:gm sharp targeted}. The detailed algorithm is shown in Algorithm.\ref{algo:sharp targeted gm}.

After the generation of poisons, we retrain the model from scratch via SGD for 160 epochs with an initial learning of 0.1 and decay by 0.1 at epochs 80,120. After training, we evaluate the prediction on targets $D_T$. For the single-target case, only predicting the target as the adversarial label is considered a success, and we sample 10 random poison-target cases as one trial, for which the average success rate over 10 cases is referred to as the success rate for this trial. For each setting(poisoning ratio and perturbation budget), we do 50 trials and report the average success rate. For the multi-target case, we randomly choose a few targets, 4 or 8 in our experiments, and the average accuracy of predicting targets as adversarial labels is referred to as the success rate of one experiment. We also repeat experiments 50 times and report the average success rate. 

\RestyleAlgo{ruled}
\begin{algorithm}[h]
\caption{\ourmodel in Targeted Attacks(Grad-Match)}
\label{algo:sharp targeted gm}
\SetKwInOut{Input}{Input}
\SetKwInOut{Output}{Output}
\Input{Pre-trained model $f(\cdot;\theta^*)$ on the clean training set $D_{tr}$; a victim set $D_T$; optimization step $M$.}
Randomly initialize the poisoning samples $D_p$\\
Compute the gradient $\nabla_{\theta}Q^S(\theta^*;D_T)$ using Eq.\ref{eq:gradient est}\\
\For{$m=1,...,M$}
 {
Update $D_p$ in Eq.\ref{eq:gm sharp targeted} with one-step gradient descent\\
}
\Output{Return $D_p$}
\end{algorithm}

\subsection{Backdoor attacks}\label{app: backdoor}

Similar to the targeted attack, we leverage the Gradient Matching method. To be more specific, we randomly choose a victim class and an objective class. Then we randomly sample the victim set $D_T$ from the victim class and select $D_p$ from the objective class to be modified in the algorithm. Note that both $D_T$ and $D_p$ are sampled from the training data. The attacking goal is to make the model retrained on $D_{tr}+D_p$ misclassify images from the victim class(in testing) which are attached with a pre-specified trigger as the objective class. We also restrict the \textit{poisoning ratio} to be 1\% and the perturbation budget is bounded by $16/255, 8/255$. 

Given a pre-trained clean model, we first randomly initialize the perturbations for $D_p$, and during each optimization step, we compute the gradient in Eq.\ref{eq:gradient est}, but different from targeted attacks, in Eq.\ref{eq:gradient est} $D_T$ are attached with the trigger which will be used to backdoor images in the testing. Then we do one-step gradient descent to minimize the similarity loss defined in Eq.\ref{eq:gm sharp targeted}, after optimizing for $R$ steps, we update the model $f$ on poisoned training data $D_{tr}+D_p$, and then continue optimizing perturbations on the updated model. The detailed algorithm is shown in Algorithm.\ref{algo:sharp backdoor gm}.

After the generation, we retrain the model from scratch via SGD for 160 epochs with an initial learning of 0.1 and decay by 0.1 at epochs 80,120. After that, we evaluate the prediction of triggered images from the entire victim class in the testing. The average accuracy of predicting triggered images as the adversarial label is referred to as the success rate. We repeat experiments 50 times and report the average accuracy as the final results. We conduct experiments on multiple datasets including CIFAR10, CIFAR100 and SVHN, along with multiple victim models including ResNet18, ResNet50 and VGG11.

\RestyleAlgo{ruled}
\begin{algorithm}
\caption{\ourmodel in Backdoor Attacks}
\label{algo:sharp backdoor gm}
\SetKwInOut{Input}{Input}
\SetKwInOut{Output}{Output}
\Input{Pre-trained model $f(\cdot;\theta^*)$ on the clean training set $D_{tr}$; a victim set $D_T$; retraining factor $R$; optimization step $M$.}
Randomly initialize the poisoning samples $D_p$\\
\For{$m=0,...,M-1$}
 {
\If{$m$ mod $\lfloor M/(R+1) \rfloor=0$ and $m\ne M$}
{
 Update $\theta^*$ on poisoned training $D_{tr}+ D_p$\\
Find the worst-case direction $\hat{v}$ using Eq.\ref{eq:estimate_v}\\
Approximate $\nabla_{\theta}^S Q(\theta^*;D_T)$ by  $\nabla_{\theta}Q(\theta^*;D_T)|_{\theta = \theta^* + \hat{v}}$.\\
}
Update $D_p$ with one step gradient descent\\
}
\Output{Poisoning set $D_p$}
\end{algorithm}

\subsection{Un-targeted attacks}\label{app: untarget}

We implement SAPA based on two existing methods: Error-min\citep{huang2021unlearnable} and Error-max\citep{fowl2021adversarial}.

\textbf{Error-min+SAPA}.
We have discussed this method in Section~\ref{sec:our_method2}, and the detailed algorithm is shown in Algorithm.\ref{algo:aa1}. In our experiments, we set $T=20$, $M=10$, $\alpha=\epsilon/10$ and $E=100$.

\textbf{Error-max+SAPA}. Different from Error-min, Error-max is used to solve the optimization as follows:
$$
\max_{\{\delta_i\}_{i=1, ...,n}} ~\sum_{(x_i,y_i)\in D_{tr}} l\big(f(x_i + \delta_i;\theta^*), y_i)
$$
where $\theta^*$ denote a model trained on clean data and is fixed during poison generation.
Thus SAPA+Error-max focuses on the sharpness-aware objective as follows:
\begin{equation*}\small
    \max_{\{\delta_i\}_{i=1, ...,n}} ~\Big[ \max_{||v||\leq \rho}\sum_{(x_i,y_i)\in D_{tr}} l\big(f(x_i + \delta_i;\theta^*+v), y_i\big)\Big]
\end{equation*}

Similar to Error-Min+SAPA, we also solve this optimization problem using gradient descent. As shown in Algorithm~\ref{algo:sharp untar max}, given the clean pre-trained model $f(\cdot,\theta)$, we first find the worst direction $\hat{v}$ to maximize  $\max_{v} ~\sum_{(x_i,y_i)\in D_{tr}} l\big(f(x_i + \delta_i;\theta^*+v), y_i)$ and then update $\delta_i$ fixing $\theta^*+\hat{v}$.

After generating perturbations via either SAPA+Error-min or SAPA+Error-max, we retrain the model on poisoned training via SGD for 160 epochs with an initial learning of 0.1 and decay by 0.1 at epochs 80,120. Then we evaluate the prediction of clean testing and report the average accuracy. Our experiment is conducted on multiple datasets including CIFAR10, CIFAR100, SVHN, along with multiple victim models including ResNet18 and ResNet50.

\begin{algorithm}[h]
\caption{Error-max+SAPA}
\label{algo:sharp untar max}
\SetKwInOut{Input}{Input}
\SetKwInOut{Output}{Output}
\Input{Network $f(\cdot;\theta)$; clean training set $\{(x_i,y_i)\}_{i=1}^n$; perturbation bound $\epsilon$; PGD step $T$; pre-train steps $R$}
\Output{Sample-wise perturbation $D_p=\{\delta_i\}_{i=1}^n$}
\For{r in $1,...,R$}{
Update $\theta$ via minimizing $L(\theta;D_{tr})$
}
Randomly initialize perturbation $D_p$\\

Fix $\theta, D_p$, find the worst direction $\hat{v}$ to maximize  $~\sum_{(x_i,y_i)\in D_{tr}} l\big(f(x_i + \delta_i;\theta^*+v), y_i)$\\
\For{$t=1,...,T$ PGD steps}{Update $D_p$ via gradient ascent fixing $\theta+\hat{v}$}

\end{algorithm}

\newpage
\section{Additional Experiments}\label{app:exp}
\vspace{-0.25cm}

In this section, we provide additional experiments to further illustrate the effectiveness of \ourmodel. 

\subsection{Additional dataset}

We test on the additional dataset, SVHN, to further illustrate the advantage of our method. All the experiments are conducted following the same procedure as in Section.\ref{sec:exp}, while some details are different. For un-targeted attacks, we repeat all experiments on SVHN. For targeted attacks, we conduct on the ``Single-Victim'' case and omit baselines Poison-Frog and Bulleyes because they do not perform well under the train-from-scratch setting. For backdoor attacks, we also omit baselines Clean-label and Hidden-trigger which do not perform well.

\textbf{Performance comparison.} Results of un-targeted attacks are shown in Table.\ref{tab:result_untarget_svhn}. Similar to results on CIFAR10 and CIFAR100, our method outperforms nearly all baselines under all settings. Results of targeted attacks are shown in Table.\ref{tab:result_targeted_svhn}. Our method outperforms all baselines significantly especially for smaller budget sizes and poison ratios. Results of backdoor attacks are shown in Table.\ref{tab:backdoor_svhn}. Our method has better performance than baselines under all settings.

\begin{table*}[h!]
\centering
\caption{\small Test Accuracy of Models Trained on Poisoned Datasets via Un-targeted Attacks. }
\resizebox{0.6\textwidth}{!}
{
\hspace{-1cm}
\begin{tabular}{c| ccc | ccc } 
\hline
\hline
 &   \multicolumn{6}{c}{\textbf{SVHN}}\\
\hline
  &   \multicolumn{3}{c|}{\textbf{ResNet18}}   &   \multicolumn{3}{c}{\textbf{ResNet50}}\\
&100\% & 80\% &50\% &100\% & 80\% &50\%  \\
\hline
\textbf{Clean$^*$}&96.0&	96.0&	96.0&	95.9&	95.9&	95.9\\
\textbf{Separable}&8.3&	92.5&	94.3&	7.8&	90.9&	93.1 \\
\textbf{Autoregressive.}&7.2&	89.5&	92.5&	7.1&	89.1&	91.2\\
\hline
\textbf{Error-Max}&5.3&	92.9&	93.4&	5.7&	92.5&	93.7\\
\textbf{Error-Max+SAPA}&4.7&	92.3&	92.6&	5.1&91.8&	92.1\\
\hline
\textbf{Error-Min}&13.8&	92.8&	95.3&	13.7&	91.3&	94.2\\
\textbf{Error-Min+SAPA}&10.2&	91.7&	93.1&	11.4&90.4&	91.8\\
\hline\hline
\end{tabular}
}
\label{tab:result_untarget_svhn}
\end{table*}

\begin{table*}[h!]
\centering
\vspace{-0.2cm}
\caption{\small Success Rate under the ``Single-Victim'' Setting in Targeted Attacks.}
\resizebox{0.8\textwidth}{!}
{
\begin{tabular}{c |c| cc | cc | cc } 
\hline
\hline
  & &   \multicolumn{2}{c|}{$16/255$}   &   \multicolumn{2}{c|}{$8/255$}   &   \multicolumn{2}{c}{$4/255$} \\
& & $\epsilon=1\%$ & 
$\epsilon=0.2\%$  & $\epsilon=1\%$ &$\epsilon=0.2\%$ & $\epsilon=1\%$ &$\epsilon=0.2\%$ \\
\hline
\multirow{3}{*}{\textbf{SVHN}}
&\textbf{Meta-Poison}&52.1&	28.7&	35.3&	21.9&	23.8&	11.2\\
&\textbf{Grad-Match}& 62.6&	37.4&	44.2&	25.7&	31.3&	17.4\\
&\textbf{SAPA}& \blue{71.2}&	\blue{48.5}&	\blue{55.7}&	\blue{38.2}&	\blue{42.1}&	\blue{23.3}\\
\hline\hline
\end{tabular}
}
\label{tab:result_targeted_svhn}
\end{table*}

\begin{table*}[h!]
\centering
\vspace{-0.2cm}
\caption{Success Rate in Backdoor Attacks on SVHN}
\resizebox{0.75\textwidth}{!}
{
\begin{tabular}{c | cc | cc | cc } 
\hline
\hline
  &   \multicolumn{2}{c|}{ResNet18}   &   \multicolumn{2}{c|}{ResNet50}   &   \multicolumn{2}{c}{VGG11} \\
&$16/255$ & $8/255$ &$16/255$ &$8/255$ & $16/255$ &$8/255$ \\
\hline
\textbf{Sleeper Agent}&92.3&	42.6&	91.6&	34.8&	86.2&	38.1\\
\textbf{SAPA-backdoor}&\blue{97.5}&	\blue{58.3}&	\blue{95.4}&	\blue{44.1}&	\blue{91.4}&	\blue{43.7}\\
\hline\hline
\end{tabular}
}
\label{tab:backdoor_svhn}
\end{table*}

\subsection{Multi-target targeted attacks}

We also perform tests on poisoning multiple targets simultaneously. We conduct experiments on model ResNet18 and two datasets Cifar10 and Cifar100. We only compare with the most powerful baseline Grad-Match under multiple settings, including different number of victim targets(4,8), perturbation size(16/255, 8/255 and 4/255) and poisoning rate($1\%$ and $0.25\%$). All results are shown in Table~\ref{tab:result_targeted_2}. It is obvious that \ourmodel consistently improves the performance under all settings.

\begin{table*}[h!]
\centering
\caption{Avg.~Success Rate under the ``Multiple-Victim'' Setting in Targeted Attacks.}
\resizebox{0.75\textwidth}{!}
{
\begin{tabular}{c|c | cc | cc | cc } 
\hline
\hline
  & &   \multicolumn{2}{c|}{$16/255$}   &   \multicolumn{2}{c|}{$8/255$}   &   \multicolumn{2}{c}{$4/255$} \\
& &1\% & 0.25\% &1\% &0.25\% & 1\% &0.25\% \\
\hline
\multirow{2}{*}{\textbf{CIFAR10: 4 Victims}}
&\textbf{Grad-Match}& 62.9&36.2&		34.7&25.1&	20.3&		7.5\\
&\textbf{SAPA}& \blue{75.1} & \blue{53.4} & \blue{47.9} &	\blue{30.8} & \blue{24.3} &	\blue{10.8} \\
\hline
\multirow{2}{*}{\textbf{CIFAR10: 8 Victims}}
&\textbf{Grad-Match}& 52.1&23.2&	27.9&	18.4&	12.7&	5.6\\
&\textbf{SAPA}& \blue{64.6} & \blue{31.2} & \blue{34.6} & \blue{26.1} & \blue{17.5} & \blue{7.3}	\\
\hline
\multirow{2}{*}{\textbf{CIFAR100: 4 Victims}}
&\textbf{Grad-Match}&	67.3&30.1	&	37.2&12.5&	17.8&2.9\\
&\textbf{SAPA}&\blue{74.2}&\blue{43.8}&\blue{44.3}&\blue{19.7}&	\blue{25.1}&\blue{6.1} \\
\hline
\multirow{2}{*}{\textbf{CIFAR100: 8 Victims}}
&\textbf{Grad-Match}&	43.6&23.2	&	16.7&4.9	&	13.8&2.7 \\ 
&\textbf{SAPA}&\blue{52.7}&\blue{31.3}&\blue{24.8}&\blue{8.3}&\blue{18.7}&\blue{4.2} \\
\hline
\end{tabular}
}
\label{tab:result_targeted_2}
\end{table*}

\subsection{Hyperparameter}

The radius $\rho$ in Eq.~\ref{eq:sharp_targeted} is the most important hyperparameter in our proposed method, and it represents the radius within which the locally worst model is searched. In the work~\citep{foret2020sharpness}, the default value is $0.05$ and we adopt it in our main experiments. However, we are interested in whether this hyperparameter has a large impact on the performance. Therefore, we test for different values, i.e. $\rho=0.01,0.05,0.1,0.2,0.5$. To avoid the computational cost, we only test on targeted(SAPA+Grad-Match) and backdoor attack(SAPA+SleeperAgent), and results are shown in Tabel~\ref{exp:hyper}. We notice that a smaller $\rho$ causes the algorithm to regress to the backbone attacks without SAPA, so the poisoning effect drops; while a very large value ($\rho\ge 0.2$) also leads to poor attack performance. 

\begin{table*}[h]
\centering
\caption{Hyperparameter $\rho$}
\label{exp:hyper}
\resizebox{0.6\textwidth}{!}
{
\begin{tabular}{c | c | c|c |c|c } 
\hline
\hline
\textbf{Method/$\rho$}&	0.01&0.05&0.1&0.2&0.5\\
\hline
\textbf{SAPA+GM}&	76.5&80.4&82.1&77.5&69.3\\
\textbf{SAPA+SleepAgent}&	92.7&97.1&97.6&96.8&83.2\\
\hline\hline
\end{tabular}
}
\end{table*}

\subsection{Ensemble and re-initialization}

In Section.\ref{sec:exp2}, we compare \ourmodel with \textit{ensemble} and \textit{re-initialization} techniques. It is also worth noting that \ourmodel can leverage these techniques to further improve the performance. We conduct experiments on targeted(SAPA+Grad-Match) and backdoor(SAPA+SleeperAgent) attacks to show this. Same as in Section.\ref{sec:exp2}, let $K,R$ denote the number of ensembles and re-initializations respectively. 

\begin{table*}[h]
\centering
\caption{Impact of ensemble(K) and re-inistailzation(R)}
\label{exp:add_kr}
\resizebox{0.4\textwidth}{!}
{
\begin{tabular}{c | cc| cc} 
\hline
\hline

&	\multicolumn{2}{c|}{\textbf{Targeted}}&\multicolumn{2}{c|}{\textbf{Backdoor}}\\
&ASR($\uparrow$)&Time/s($\downarrow$)&ASR($\uparrow$)&Time/s($\downarrow$)\\
\hline
\textbf{K=1,R=1}&80.0&156.9&97.1&4031.3\\
\textbf{K=2,R=1}&83.5&319.1&98.2&8129.7\\
\textbf{K=4,R=1}&86.3&748.4&98.9&16254.3\\
\textbf{K=8,R=1}&87.1&1562.5&100&30888.9\\
\textbf{K=1,R=2}&81.7&296.8&97.9&8006.4\\
\textbf{K=1,R=4}&83.5&718.5&98.5&15986.7\\
\textbf{K=1,R=8}&86.2&1379.7&99.3&32459.8\\
\hline\hline
\end{tabular}
}
\end{table*}

\subsection{More results}
We also provide the performance of un-targeted attacks with standard error reported, and of targeted attacks for ResNet50 on Cifar10 in Table~\ref{exp:result_un_2} and \ref{tab:result_targeted_3} respectively. 

\begin{table*}[h]
\caption{\small Model Test Accuracy under Un-targeted Attacks. (with standard error reported)}
\label{exp:result_un_2}
\hspace{-1cm}
\resizebox{1.1\textwidth}{!}
{
\begin{tabular}{c| ccc | ccc | ccc | ccc } 
\hline
\hline
 &   \multicolumn{6}{c|}{\textbf{CIFAR10}} &   \multicolumn{6}{c}{\textbf{CIFAR100}}   \\
\hline
  &   \multicolumn{3}{c|}{\textbf{ResNet18}}   &   \multicolumn{3}{c|}{\textbf{ResNet50}}   &   \multicolumn{3}{c|}{\textbf{ResNet18}} 
 & \multicolumn{3}{c}{\textbf{ResNet50}}\\
&100\% & 80\% &50\% &100\% & 80\% &50\%  &100\% & 80\% &50\%   &100\% & 80\% &50\%  \\
\hline
\textbf{Separable} &13.5$\pm$0.11	&86.3$\pm$0.17	&92.9$\pm$0.26&14.9$\pm$0.12&88.1$\pm$0.17&93.2$\pm$0.11&9.14v0.16&57.1$\pm$0.21&66.2$\pm$0.19&8.4$\pm$0.14&60.8$\pm$0.17&66.7$\pm$0.25 \\
\textbf{Autoregressive} &11.7$\pm$0.13&82.2$\pm$0.25&89.7$\pm$0.28&10.07$\pm$0.09&83.6$\pm$0.15&90.3$\pm$0.11&4.24$\pm$0.08&51.6$\pm$0.13&64.7$\pm$0.10&4.32$\pm$0.11&53.5$\pm$0.21&66.1$\pm$0.19\\
\hline
\textbf{Error-Max}&  15.4$\pm$0.23&88.2$\pm$0.34&92.2$\pm$0.27&	27.8$\pm$0.38&90.1$\pm$0.25&93.9$\pm$0.31&4.87$\pm$0.11&57.3$\pm$0.14&66.9$\pm$0.13&5.61$\pm$0.13&58.3$\pm$0.18&68.1+0.27\\
\textbf{Error-Max+SAPA} & 11.3$\pm$0.11&84.6$\pm$0.23&90.1$\pm$0.19&	15.76$\pm$0.34&85.7$\pm$0.39&91.3$\pm$0.23&4.13$\pm$0.08&55.1$\pm$0.12&64.8$\pm$0.09&4.87$\pm$0.17&56.8$\pm$0.14&66.9$\pm$023\\
\hline
\textbf{Error-Min}& 21.2$\pm$0.26&87.1$\pm$0.38&93.4$\pm$0.35&	18.89$\pm$0.41&89.5$\pm$0.36&94.5$\pm$0.29&11.2$\pm$0.19&56.9$\pm$0.25&67.7$\pm$0.17&10.8$\pm$0.14&60.5$\pm$0.21&70.3$\pm$0.24\\
\textbf{Error-Min+SAPA}& 10.9$\pm$0.15&83.7$\pm$0.32&90.0$\pm$0.38	& 10.3$\pm$0.37&85.2$\pm$0.39&91.8$\pm$0.33&8.73$\pm$0.21&53.1$\pm$0.13&65.3$\pm$0.15&9.52$\pm$0.12&57.9$\pm$0.17&67.6$\pm$0.18\\
\hline\hline
\end{tabular}
}
\end{table*}

\begin{table*}[h!]
\centering
\vspace{-0.2cm}
\caption{\small Success Rate in Targeted Attacks(for ResNet50 on Cifar10).}
\vspace{-0.2cm}
\resizebox{0.85\textwidth}{!}
{
\begin{tabular}{c |c| cc | cc | cc } 
\hline
\hline
  & &   \multicolumn{2}{c|}{$16/255$}   &   \multicolumn{2}{c|}{$8/255$}   &   \multicolumn{2}{c}{$4/255$} \\
& & $\epsilon=1\%$ & 
$\epsilon=0.2\%$  & $\epsilon=1\%$ &$\epsilon=0.2\%$ & $\epsilon=1\%$ &$\epsilon=0.2\%$ \\
\hline
\multirow{4}{*}{\textbf{CIFAR10}} &\textbf{Bullseye}&1.7$\pm$0.9&0.5$\pm$0.2&1.1$\pm$0.6&0.4$\pm$0.1&0$\pm$0&0$\pm$0 \\
&\textbf{Poison-Frog}&2.3$\pm$0.7&0.6$\pm$0.4&0.8$\pm$0.3&0$\pm$0&0$\pm$0&0$\pm$0\\
&\textbf{Meta-Poison}&46.9$\pm$2.5&34.1$\pm$2.6&23.5$\pm$1.7&10.7$\pm$1.6&15.6$\pm$1.3&6.9$\pm$1.2\\
&\textbf{Grad-Match}&75.6$\pm$3.7&53.9$\pm$4.3&30.1$\pm$3.8&14.8$\pm$2.6&21.5$\pm$2.9&9.6$\pm$2.3\\
&\textbf{SAPA}&\blue{81.4$\pm$3.1}&\blue{60.2$\pm$3.5}&\blue{34.7$\pm$3.6}&\blue{17.6$\pm$1.9}&\blue{26.3$\pm$2.1}&\blue{12.4$\pm$1.9}\\
\hline
\multirow{4}{*}{\textbf{CIFAR100}} &\textbf{Bullseye}&2.4$\pm$1.1&1.2$\pm$0.5&1.6$\pm$0.9&0.7$\pm$0.3&0$\pm$0&0$\pm$0\\
&\textbf{Poison-Frog}&2.9$\pm$0.9&1.3$\pm$0.4&1.5$\pm$0.7&0.8$\pm$0.2&0$\pm$0&0$\pm$0\\
&\textbf{Meta-Poison}&50.2$\pm$2.4&29.8$\pm$1.8&27.9$\pm$2.1&12.5$\pm$1.1&19.4$\pm$1.6&8.2+1.8\\
&\textbf{Grad-Match}&80.8$\pm$3.6&50.7$\pm$4.1&36.7$\pm$3.8&24.3$\pm$3.9&26.9$\pm$2.7&10.4$\pm$2.1\\
&\textbf{SAPA}&\blue{84.5$\pm$3.4}&\blue{56.4$\pm$2.9}&\blue{42.8$\pm$3.4}&\blue{29.6$\pm$2.7}&\blue{30.1$\pm$2.4}&\blue{13.5$\pm$1.6}\\
\hline\hline
\end{tabular}
}
\label{tab:result_targeted_3}
\end{table*}

\section{Additional visualizations}

\begin{figure}[h!]
\subfloat[\footnotesize Backdoor(SleeperAgent)]{ 
\begin{minipage}[c]{0.25\textwidth}
\centering
\includegraphics[width = 1\textwidth]{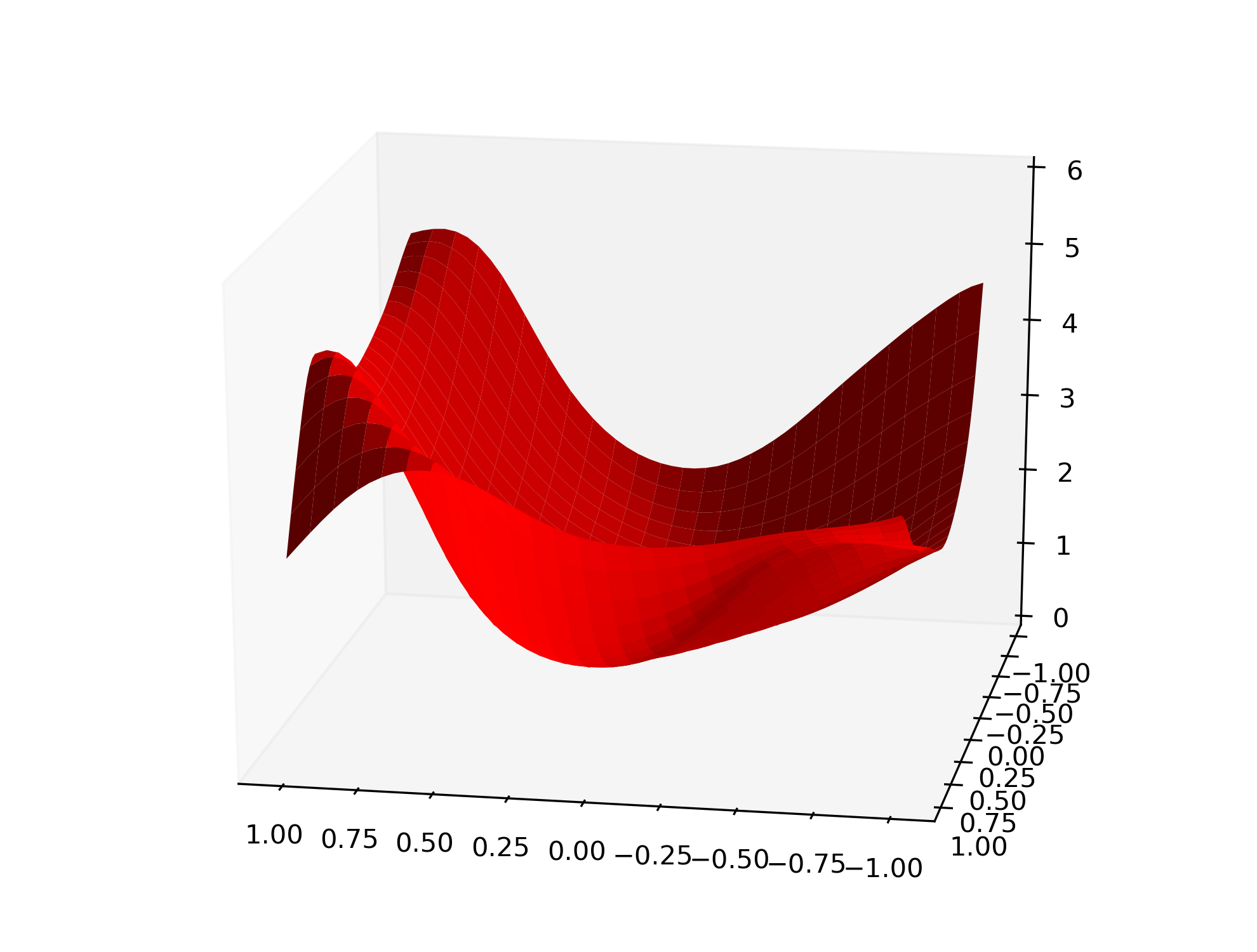}
\end{minipage}
}
\subfloat[\footnotesize Backdoor(SAPA)]{ 
\begin{minipage}[c]{0.25\textwidth}
\centering
\includegraphics[width = 1\textwidth]{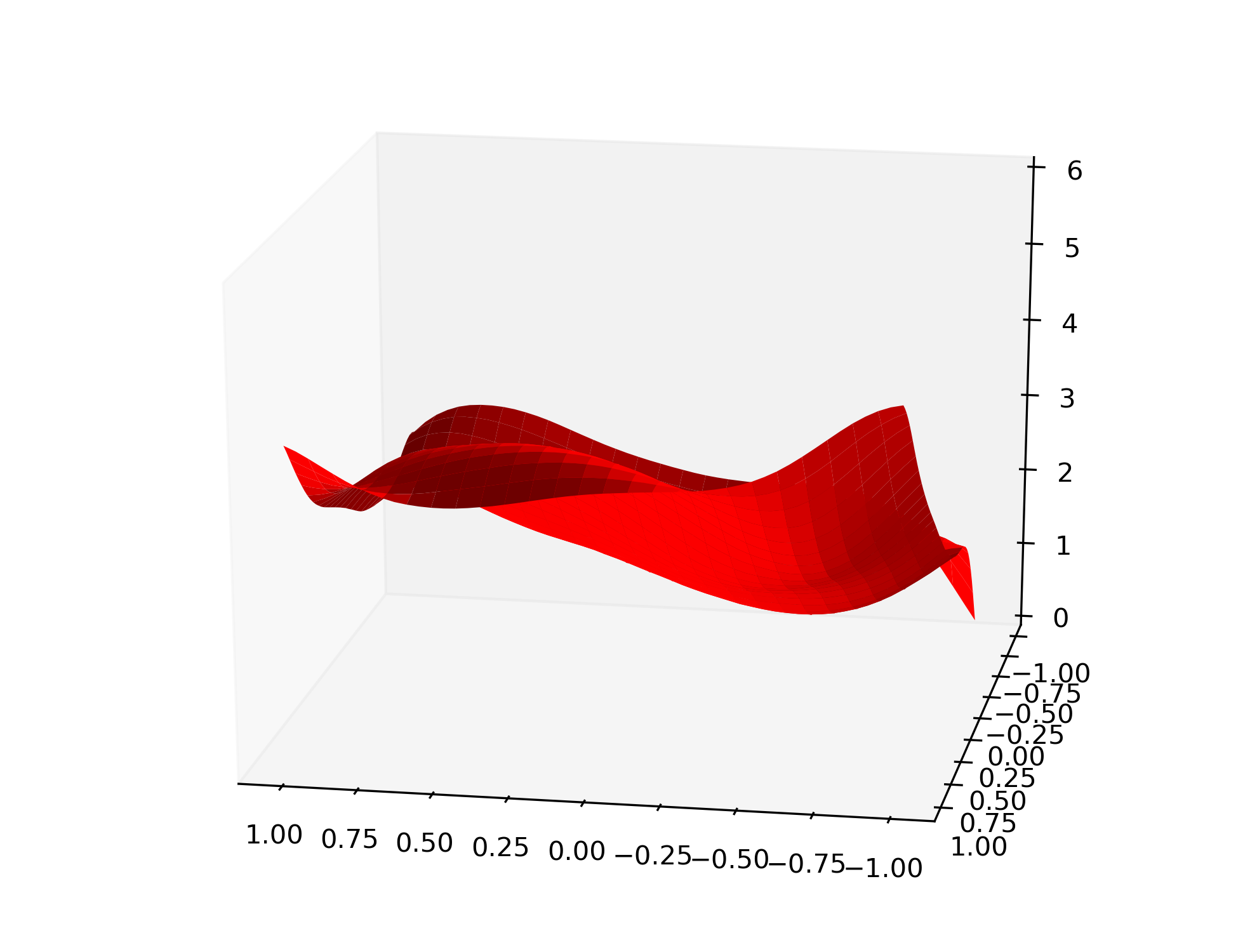}
\end{minipage}
}
\subfloat[\footnotesize Targeted(Grad-Match)]{ 
\begin{minipage}[c]{0.25\textwidth}
\centering
\includegraphics[width = 1\textwidth]{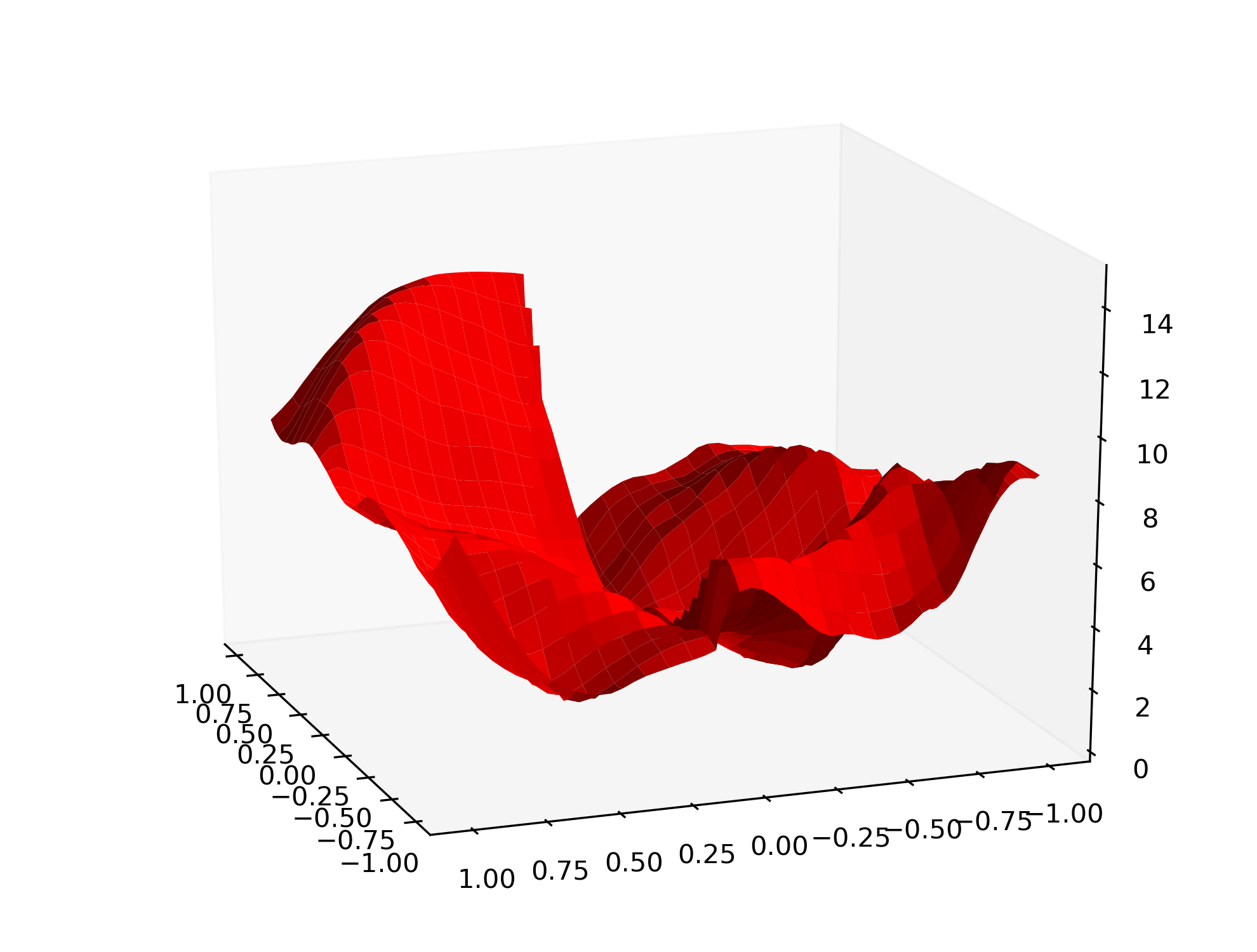}
\end{minipage}
}
\subfloat[\footnotesize Targeted(SAPA)]{ 
\begin{minipage}[c]{0.25\textwidth}
\centering
\includegraphics[width = 1\textwidth]{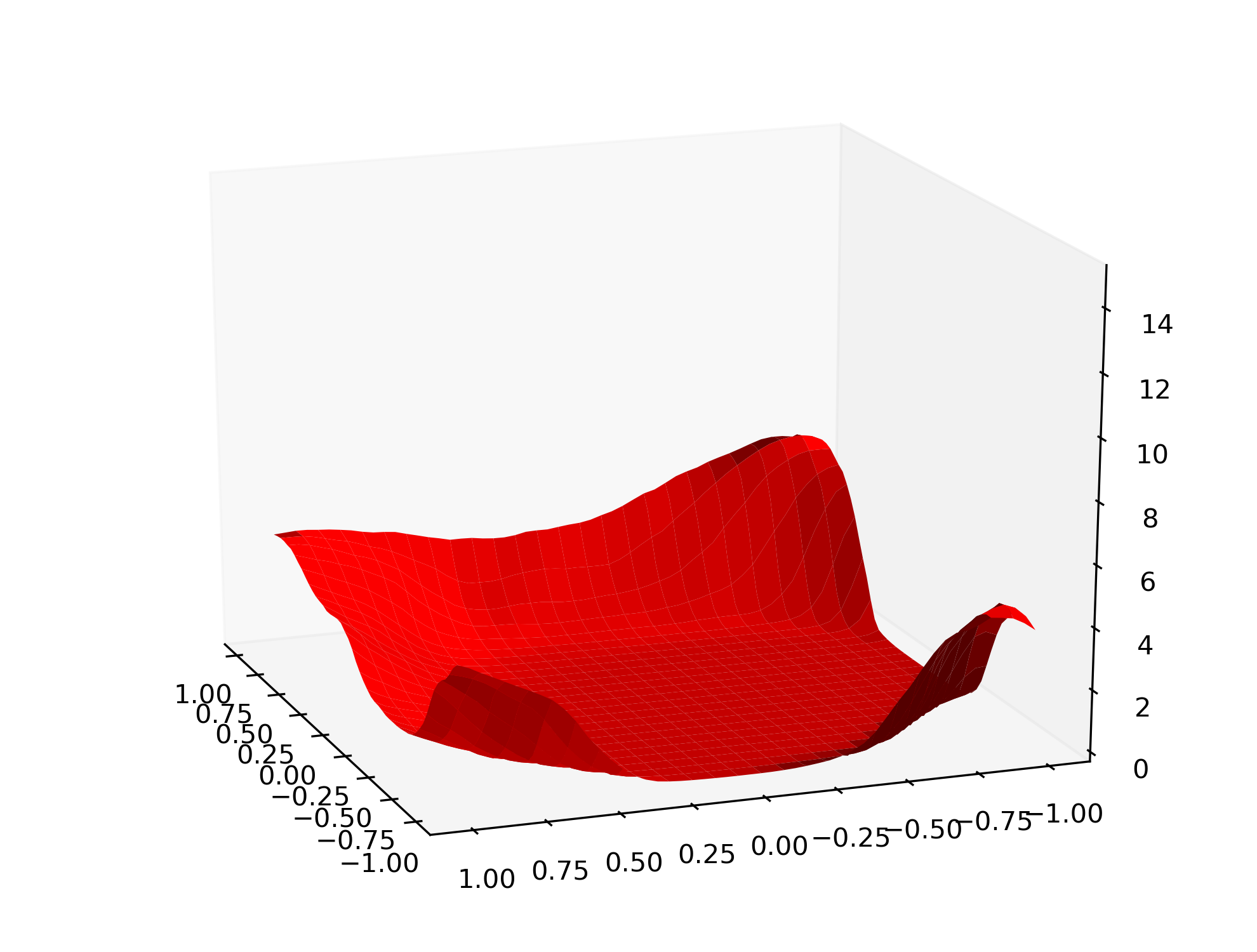}
\end{minipage}
}
\caption{\small Visualization of loss landscape for Backdoor and Targeted attacks.}
\label{fig:add_visual}
\end{figure}

\textbf{Loss Landscape}.
We provide the visualization of the loss landscape for targeted and backdoor attacks to further illustrate that our method can indeed minimize the sharpness. For both attacks, losses are computed on the victim set $D_T$, and we visualize on CIFAR10 and ResNet18 with perturbation size $16/255$ and poison ration 1\%. Visualizations are shown in Figure.\ref{fig:add_visual}. It is obvious that our method can achieve a smoother loss landscape.

\begin{figure}[h!]
\centering
\subfloat[\small 40 epochs, $16/255$]{\label{fig:ad2}
\begin{minipage}[c]{0.25\textwidth}
\centering
\includegraphics[width = 1\textwidth]{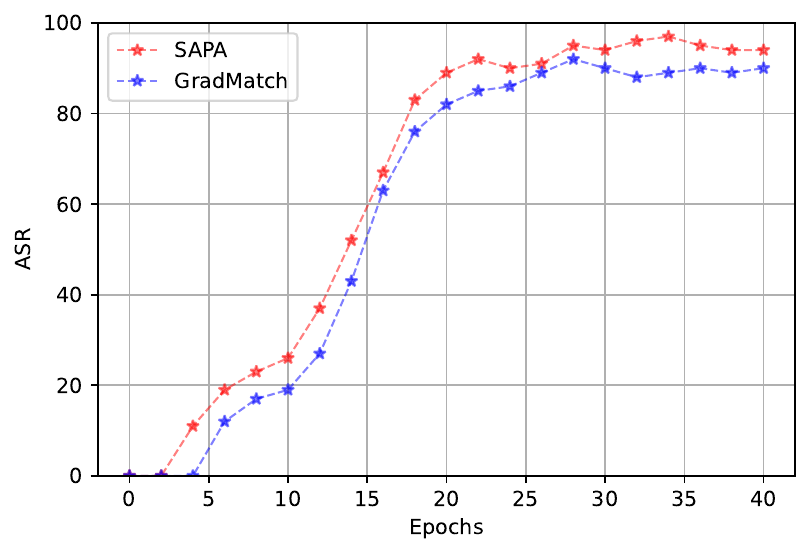}
\end{minipage}
}
\subfloat[\small 160 epochs, $16/255$]{\label{fig:ad1}
\begin{minipage}[c]{0.25\textwidth}
\centering
\includegraphics[width = 1\textwidth]{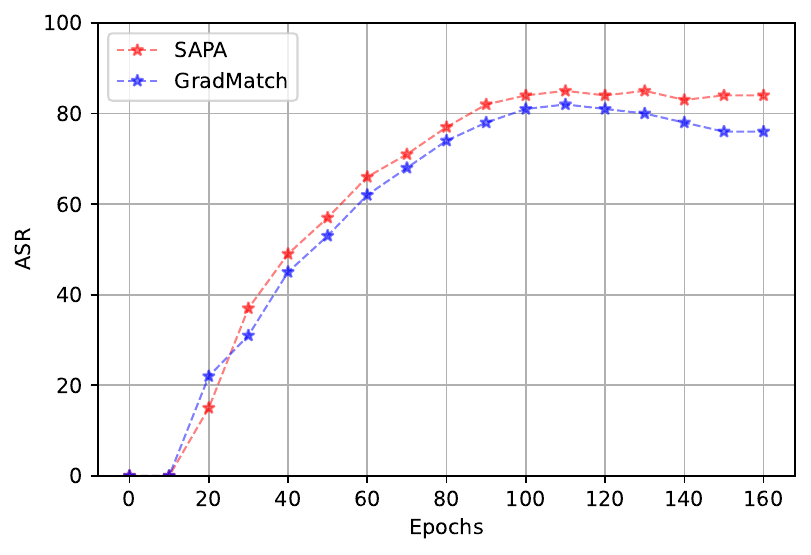}
\end{minipage}
}
\subfloat[\small 40 epochs, $8/255$]{\label{fig:ad3}
\begin{minipage}[c]{0.25\textwidth}
\centering
\includegraphics[width = 1\textwidth]{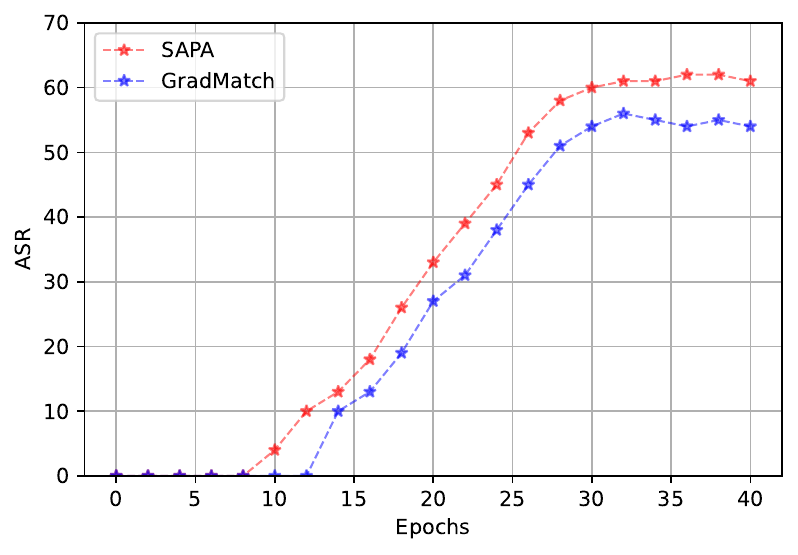}
\end{minipage}
}
\subfloat[\small 160 epochs, $8/255$]{\label{fig:ad4}
\begin{minipage}[c]{0.25\textwidth}
\centering
\includegraphics[width = 1\textwidth]{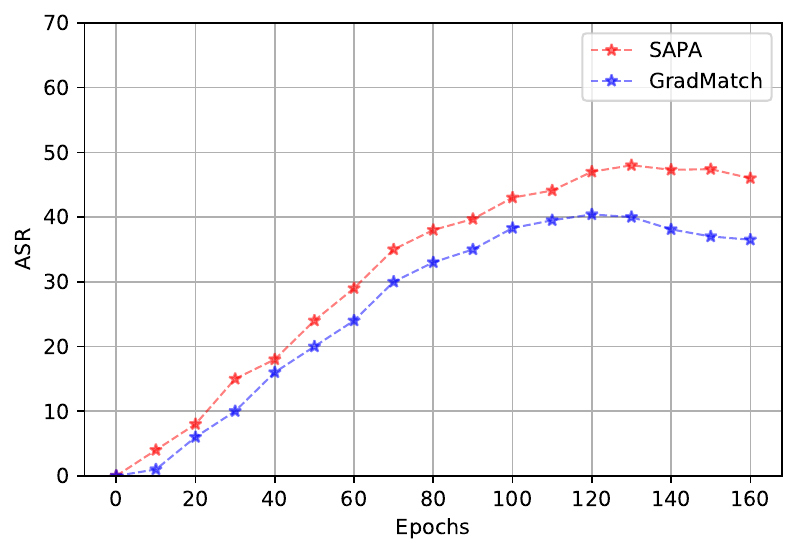}
\end{minipage}
}
\caption{\small Attack successful rate (ASR) for different training epochs of targeted attacks}
\label{fig:epochs}
\end{figure}

\textbf{Effect of epochs on targeted}. As we mentioned in Section~\ref{sec:exp}, we adopt different re-training epochs when applying baseline Grad-Match. In specific, the original paper\citep{geiping2020witches} adapts 40 epochs while we re-trained for 150 epochs. We plot the success rates for different epochs(40 and 160) under different perturbation sizes(16/255 and 8/255) and show them in Figure~\ref{fig:epochs}. Results of \ourmodel and Grad-Match are in red and blue respectively. From the figures, we notice that the 160-epoch has a smaller success rate than the 40-epoch and the success rate is slightly decreasing when the epochs are growing. These observations imply that re-training epochs indeed influence the poisoning effect and \ourmodel can improve the performance to some extent.

\end{document}